\documentclass[preprint]{aastex63}

\newcommand\rrle{RR Lyr\ae}
\newcommand\rrl{RRL}
\newcommand\rrls{RRLs}

\defcitealias{Martinez-Vazquez2016}{MV16}
\defcitealias{2013ApJ...773..181N}{N13}
\defcitealias{1996A&A...312..111J}{JK96}
\defcitealias{ZinnWest1984}{ZW84}
\defcitealias{2019ApJ...882..169F}{F19}
\defcitealias{2020arXiv201202284C}{C20}
\defcitealias{Carretta2009}{C09}
\defcitealias{2020arXiv200802280I}{IB20}

\received{October 23, 2020}
\revised{March 16, 2021}
\accepted{March 16, 2021}
\submitjournal{ApJ}

\shorttitle{\rrl{} Fourier Metallicity from Light Curves}
\shortauthors{Mullen et al.}

\begin{document}

\title{Metallicity of Galactic \rrle{} from Optical and Infrared Light Curves:\\ I. Period-Fourier-Metallicity Relations for Fundamental Mode \rrle{}}

\correspondingauthor{Joseph P. Mullen}
\email{jpmullen@iastate.edu}

\author[0000-0002-1650-2764]{Joseph P. Mullen}
\affiliation{Department of Physics and Astronomy, Iowa State University, Ames, IA 50011, USA}

\author[0000-0001-9910-9230]{Massimo Marengo}
\affiliation{Department of Physics and Astronomy, Iowa State University, Ames, IA 50011, USA}

\author[0000-0002-9144-7726]{Clara E. Mart\'inez-V\'azquez}
\affiliation{Cerro Tololo Inter--American Observatory, NSF’s National Optical-Infrared Astronomy Research Laboratory, Casilla 603, La Serena, Chile}

\author[0000-0002-8894-836X]{Jillian R. Neeley}
\affiliation{Department of Physics, Florida Atlantic University, 777 Glades Rd, Boca Raton, FL 33431}

\author[0000-0002-4896-8841]{Giuseppe Bono}
\affiliation{Dipartimento di Fisica, Universit\`a di Roma Tor Vergata, via della Ricerca Scientifica 1, 00133 Roma, Italy}
\affiliation{INAF -- Osservatorio Astronomico di Roma, via Frascati 33, 00078 Monte Porzio Catone, Italy}

\author[0000-0001-8209-0449]{Massimo Dall'Ora}
\affiliation{INAF -- Osservatorio Astronomico di Capodimonte, Salita Moiariello 16, 80131 Napoli, Italy}

\author[0000-0003-3096-4161]{Brian Chaboyer}
\affiliation{Department of Physics and Astronomy, Dartmouth College, 6127 Wilder Laboratory, Hanover, NH 03755, USA}

\author[0000-0002-5032-2476]{Fr\'ed\'eric Th\'evenin}
\affiliation{Universit\'e de Nice Sophia--antipolis, CNRS, Observatoire de la C\^ote d'Azur, Laboratoire Lagrange, BP 4229, F-06304 Nice, France}

\author[0000-0001-7511-2830]{Vittorio F. Braga}
\affiliation{INAF -- Osservatorio Astronomico di Roma, via Frascati 33, 00078 Monte Porzio Catone, Italy}
\affiliation{Space Science Data Center -- ASI, via del Politecnico snc, 00133 Roma, Italy}

\author[0000-0002-7717-9227]{Juliana Crestani}
\affiliation{Dipartimento di Fisica, Universit\`a di Roma Tor Vergata, via della Ricerca Scientifica 1, 00133 Roma, Italy}
\affiliation{INAF -- Osservatorio Astronomico di Roma, via Frascati 33, 00078 Monte Porzio Catone, Italy}
\affiliation{Departamento de Astronomia, Universidade Federal do Rio Grande do Sul, Av. Bento Gon\c{c}alves 6500, Porto Alegre 91501-970, Brazil}

\author[0000-0001-5829-111X]{Michele Fabrizio}
\affiliation{INAF -- Osservatorio Astronomico di Roma, via Frascati 33, 00078 Monte Porzio Catone, Italy}
\affiliation{Space Science Data Center -- ASI, via del Politecnico snc, 00133 Roma, Italy}

\author[0000-0003-0376-6928]{Giuliana Fiorentino}
\affiliation{INAF -- Osservatorio Astronomico di Roma, via Frascati 33, 00078 Monte Porzio Catone, Italy}

\author[0000-0003-4510-0964]{Christina K. Gilligan}
\affiliation{Department of Physics and Astronomy, Dartmouth College, 6127 Wilder Laboratory, Hanover, NH 03755, USA}

\author[0000-0001-5292-6380]{Matteo Monelli}
\affiliation{IAC- Instituto de Astrof\'isica de Canarias Calle V\'ia Lactea s/n, E-38205 La Laguna, Tenerife, Spain\\}
\affiliation{Departmento de Astrof\'isica, Universidad de La Laguna, E-38206 La Laguna, Tenerife, Spain\\}


\author[0000-0001-6074-6830]{Peter B. Stetson}
\affiliation{Herzberg Astronomy and Astrophysics, National Research Council, 5071 West Saanich Road, Victoria, British Columbia V9E 2E7, Canada}



\begin{abstract}
We present newly-calibrated period-$\phi_{31}$-[Fe/H] relations for fundamental mode \rrle{} stars in the optical and, for the first time, mid-infrared. This work’s calibration dataset provides the largest and most comprehensive span of parameter space to date with homogeneous metallicities from $-3\la \textrm{[Fe/H]}\la 0.4$ and accurate Fourier parameters derived from 1980 ASAS-SN ($V$-band) and 1083 WISE (NEOWISE extension, $W1$ and $W2$ bands) \rrle{} stars with well-sampled light curves. We compare our optical period-$\phi_{31}$-[Fe/H] with those available in the literature and demonstrate that our relation minimizes systematic trends in the lower and higher metallicity range. Moreover, a direct comparison shows that our optical photometric metallicities are consistent with both those from high-resolution spectroscopy and globular clusters, supporting the good performance of our relation. We found an intrinsic scatter in the photometric metallicities (0.41~dex in the $V$-band and 0.50~dex in the infrared) by utilizing large calibration datasets covering a broad metallicity range. This scatter becomes smaller when optical and infrared bands are used together (0.37~dex). Overall, the relations derived in this work have many potential applications, including large-area photometric surveys with JWST in the infrared and LSST in the optical.

\end{abstract}




\section{Introduction}\label{sec:intro}

RR Lyr\ae{} stars (\rrls) are the most widely used tracers of old (age $> 10$~Gyr, \citealt{Walker1989}) stellar populations in the Milky Way and Local Group galaxies (see e.g. \citealt{2015pust.book.....C} for a review). They can also be used as standard candles thanks to a well defined $M_V$ vs. iron abundance relation \citep{1990ApJ...350..603S, 1998A&ARv...9...33C}. The recent calibration of theoretical \citep{2015ApJ...808...50M, 2017ApJ...841...84N} and observational \citep{2013MNRAS.435.3206D, 2018MNRAS.481.1195M, 2019MNRAS.490.4254N}, period-Wesenheit and period-luminosity relations (in the optical and infrared, respectively) have revealed the true potential of these stars as high precision distance indicators. These same studies have highlighted the role of metallicity in determining the absolute brightness of these variables, hence requiring period-luminosity-metallicity (PLZ) and period-Wesenheit-metallicity (PWZ) relations for a reliable estimate of their distances. Accurate measurements of \rrl{} metal abundances, however, are hard to come by.

Nearly all currently available catalogs (see e.g. \citealt{2013MNRAS.435.3206D} for a recent large compilation) tend to list [Fe/H] abundances derived with  heterogeneous methods and often calibrated with different scales. Metallicities derived from high-resolution spectra ($R \ga 20$,000) offer the highest level of precision ($\sim 0.1$~dex, e.g. \citealt{1995AJ....110.2319C,1996ApJS..103..183L,1996A&A...312..957F,2011ApJS..197...29F, 2013ApJ...773..181N, 2014ApJ...782...59G, 2015MNRAS.447.2404P, 2017ApJ...835..187C,2018ApJ...864...57M, 2019ApJ...881..104M} and Gilligan et al., submitted), but require large amounts of telescope and analysis time, and therefore exist only for a small number of stars. Medium-resolution spectroscopic surveys ($R \sim 2$,000), however, are available for larger samples of \rrls. For example, over 2,300 sources found in the Sloan Extension for Galactic Understanding and Exploration (SEGUE) survey \citep{2009AJ....137.4377Y}, and another $\sim 3$,000 targets from the Large Scale Area Multi-Object Spectroscopic Telescope (LAMOST) DR2 survey \citep{2012RAA....12..735D, 2014IAUS..298..310L}, can also be leveraged to derive iron abundances. This is traditionally achieved with the $\Delta S$ method \citep{1959ApJ...130..507P}, which relies on ratios between the equivalent widths of Ca and H lines. Recent applications of this method are shown in both \citet{2020ApJS..247...68L}, which is based on a fit of theoretical models and utilizes LAMOST data, and \citet{2020arXiv201202284C} (C20 hereafter), which is based upon empirical calibrators and is shown to be valid for both SEGUE and LAMOST data. These applications demonstrate the viability of the $\Delta S$ method in providing homogeneous [Fe/H] abundances with an uncertainty of 0.2-0.3~dex.

While the $\Delta S$ method allows a dramatic expansion to the sample of \rrls{} with known metallicity to thousands of sources, it still requires spectra. As such, it is not directly applicable to the much larger number of \rrls{} that will be discovered within the Milky Way and the other Local Group galaxies by upcoming large area photometric surveys such as the Rubin observatory Legacy Survey of Space and Time (LSST) \citep{2019ApJ...873..111I} in the optical, and surveys that will be executed for the Roman telescope \citep{2019arXiv190205569A} at near-infrared wavelength. A reliable and precise method to derive metallicities from photometric time series is necessary to enable an expansion of such measurements to distances where taking spectra is not possible at all, or in high extinction environments that can only be probed photometrically by mid-infrared telescopes such as the James Webb Space Telescope (JWST) \citep{2006SSRv..123..485G}.

Early work by \citet{1996A&A...312..111J} (JK96 hereafter) has demonstrated that at optical wavelengths the shape of the light curve of fundamental mode \rrls{} (RRab) is related to their metallic abundance. In particular, they derived a linear relation connecting RRab's [Fe/H] abundance with period and low order parameters in the Fourier decomposition of the star's light curve in the $V$-band, with the phase parameter $\phi_{31} = \phi_3 - 3 \cdot \phi_1$ providing the most sensitive diagnostics. Further work by \citet{2013ApJ...773..181N} (hereafter N13), \citet{Martinez-Vazquez2016} (hereafter MV16), \citet{2005AcA....55...59S}, \citet{2016ApJS..227...30N}, and \citet{2020arXiv200802280I} (hereafter IB20) extended this analysis, respectively, to well-sampled RRab light curves obtained with the Kepler space telescope \citep{2010ApJ...713L..79K}, to include stars in globular clusters, to the Optical Gravitational Lensing Experiment (OGLE, \citealt{1992AcA....42..253U}) $I$-band, to the Palomar Transient Factory (PTF, \citealt{2009PASP..121.1395L}) $R$-band, and by using Gaia DR2 \citep{2018A&A...616A...1G,2018A&A...618A..30H,2019A&A...622A..60C} $G$-band light curves.

In this work, we take advantage of both the large sample of homogeneous HR [Fe/H] abundances in \citetalias{2020arXiv201202284C} and apply the $\Delta$S calibration of \citetalias{2020arXiv201202284C} to the full medium-resolution LAMOST DR6 and SDSS-SEGUE datasets to build an extensive HR+$\Delta$S metallicity catalog. We have cross-matched the variables in the HR+$\Delta$S metallicity catalog with well-sampled photometric time series in the All-Sky Automated Survey for Supernovae (ASAS-SN,  \citealt{2014ApJ...788...48S, 2018MNRAS.477.3145J}) and the Near-Earth Objects reactivation mission (NEOWISE, \citealt{2011ApJ...731...53M}) of the  Wide-field Infrared Survey Explorer (WISE, \citealt{2010AJ....140.1868W}). We have then derived novel period-$\phi_{31}$-[Fe/H] relations in the optical ($V$-band) and, for the first time, mid-infrared ($W1$ and $W2$ bands). Our work shows that these relations can indeed be extended to the thermal infrared, where the light curves are mostly determined by the radius variation during the star's pulsation rather than the effective temperature changes that dominate in the optical wavelengths. As mentioned above, this will be crucial to allow the determination of reliable metallicities in upcoming space infrared surveys.

This paper is structured as follows. In Section~\ref{sec:data}, we describe in detail the data sets we adopt for our work: the HR+$\Delta$S metallicity catalog utilizing the work of \citetalias{2020arXiv201202284C}, the ASAS-SN and WISE time-series catalogs, and the light curves for a sample of Galactic globular clusters with known metallicity that we will use to validate our relations. In Section~\ref{sec:fourier}, we explain how our period-$\phi_{31}$-[Fe/H] relations are calibrated and validated. Our results are discussed in Section~\ref{sec:discussion}, where we assess the precision of the infrared and optical relations, compare our relations with previous ones found in literature, and apply our method to measure the [Fe/H] abundance in the sample of Galactic globular clusters. Our conclusions are presented in Section~\ref{sec:conclusions}.


\begin{table}[!h]

    \begin{center}
    \caption{Calibration Datasets}\label{tab:Calibrators}
    \begin{tabular}{lcccc}
    \tableline
    & ASAS-SN & \multicolumn{2}{c}{WISE}& Joint sample \\
    \tableline
    \tableline
    Bands & $V$ & $W1$ & $W2$ & $V$, ($W1$ or $W2$)  \\
    RRab stars & 1980 & 1083 & 707 & 967 \\
    Period in days (range) &0.36 - 0.89 &0.36 - 0.85 &0.36 - 0.85 &0.36 - 0.85 \\
    Period in days (mean value)&0.57 &0.57 &0.57 &0.57\\
    $[$Fe/H$]$ (range) &$-$3.06 - (+0.36) &$-$3.06 - (+0.36) &$-$3.06 - (+0.36) & $-$3.06 - (+0.36) \\
    $[$Fe/H$]$ (mean value) &$-$1.47 &$-$1.46 &$-$1.44 &$-$1.45 \\
    Number of epochs (range)\tablenotemark{a} &69 - 892 &153 - 879 &153 - 879 & \nodata \\
    Number of epochs (mean value) &270 &231 &233 & \nodata \\
    Magnitude (range) &9.55 - 17.41 &7.85 - 14.76 &7.87 - 14.11 & \nodata \\ 
    \tableline
    \end{tabular}  
    \end{center}

    \tablenotetext{a}{The distribution of epochs is recorded prior to removing any spurious photometric measurement, as described in Section~\ref{sec:fourier}.}
\end{table}

\section{Field and Globular Cluster Fundamental RR Lyrae Datasets} \label{sec:data}

In this section, we describe the properties of the sample of \rrls{} with known spectroscopic [Fe/H] abundances that we have adopted to calibrate our period-$\phi_{31}$-metallicity relations. From this catalog, we have derived three \emph{calibration samples}: one with $V$-band time-series (\emph{ASAS-SN sample}), one with photometric data available in the thermal infrared (\emph{WISE sample}), and one with data in both wavelengths (\emph{joint sample}). The properties of these calibration samples are described in the following sections and listed in Table~\ref{tab:Calibrators}.


\subsection{Calibration Sample} \label{ssec:sample}
As mentioned before, the sources selected for this analysis are chosen from both the high-resolution metallicity catalog of \citetalias{2020arXiv201202284C} and the full medium-resolution LAMOST DR6 and SDSS-SEGUE datasets, from which the spectrum selection criteria and $\Delta$S metallicity calibration of \citetalias{2020arXiv201202284C} have been applied. The resultant HR+$\Delta$S metallicity catalog is comprised of 8660 fundamental mode field \rrl{}s that have also been cross-matched with the Gaia EDR3 database \citep{gaiacollaboration2020gaia}, as well as a number of other publicly available datasets, using an algorithm specifically developed for sparse catalogs \citep{2019A&A...621A.144M}.

The HR+$\Delta$S metallicity catalog provides both a homogenized sample of RRab iron abundances gathered from various sources in literature (170 of which are derived from high-resolution spectra) and similarly homogeneous new $\Delta S$ metallicity estimates, as the $\Delta S$ calibration of \citetalias{2020arXiv201202284C} is based in part on the metallicity of 111 of the aforementioned HR RRab stars. For a complete and detailed description of the HR metallicity catalog's demographics, the $\Delta S$ calibration, and the spectrum selection criterion, we refer the reader to the \citetalias{2020arXiv201202284C} paper.

All metallicities in this catalog are based upon the metallicity scale utilized by \citetalias{2020arXiv201202284C}, which is based upon high-resolution spectroscopy, and the most updated iron line parameters, most of which have transition parameters derived in laboratory studies. This is the same scale utilized by \citet{2011ApJS..197...29F}, \citet{2017ApJ...835..187C}, and \citet{Sneden_2017}, and will be the default scale used throughout this paper unless otherwise declared. Note, other literature field-\rrl{}  high-resolution works can be brought to the same metallicity scale with the addition of a simple offset. A full analysis of the offsets between various HR [Fe/H] scales is offered in \citetalias{2020arXiv201202284C}. It is worth mentioning here that the often used \citet[][C09]{Carretta2009} [Fe/H] scale can be converted to this works scale with the addition of a small rigid shift of 0.08 dex.

\begin{figure}[!t]
    \centering
    \includegraphics[width=\textwidth]{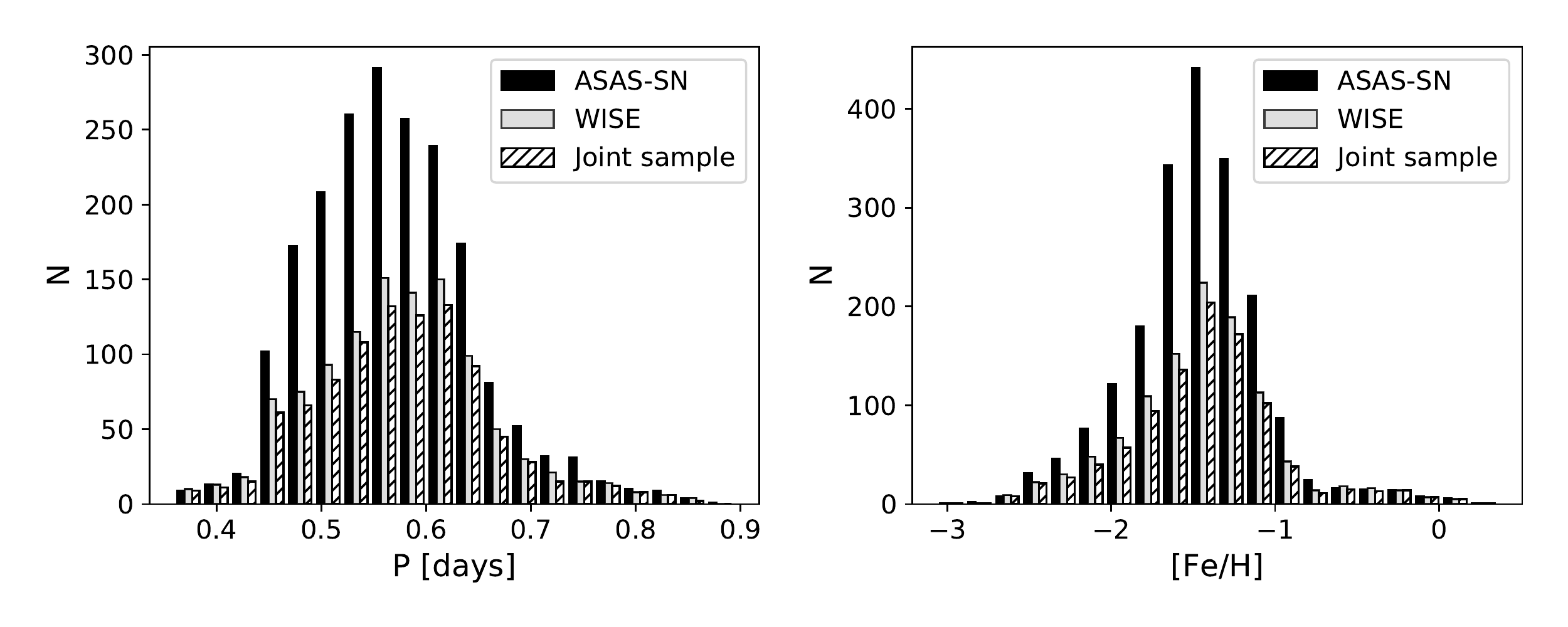}
    \caption{Period distribution (left) and spectroscopic [Fe/H] distribution (right) of the different calibration datasets. The histogram labeled ``joint sample'' (hatched) corresponds to those stars in common between the ASAS-SN (optical $V$-band, in black) and WISE (infrared $W1$, $W2$ bands, in grey) datasets.}
    \label{fig:P&Feh}
\end{figure}

In this work, we focus on the stars in the HR+$\Delta$S metallicity catalog that have a match in the ASAS-SN and WISE surveys. Figure~\ref{fig:P&Feh} shows the distribution of period and metallicity for the subset of stars with available optical (ASAS-SN) or infrared (WISE) \emph{good quality} light curves (i.e. passing the stringent photometric and Fourier decomposition criteria described in Section~\ref{sec:fourier}). The joint sample, also shown in the figure, comprises the smaller subset of \rrl{}s with light curves available at both optical and infrared wavelengths. All samples cover the entire period range expected for RRab variables and are well representative of the metallicity of Galactic Halo \rrls, with a mean [Fe/H] abundance of $\approx -1.45$ in both the ASAS-SN and WISE samples. Figure~\ref{fig:P&Feh} also shows that all histograms retain both the low metallicity $\textrm{[Fe/H]} \la -2.2$ and the high metallicity $\textrm{[Fe/H]} \ga -0.7$ tail present in the \citetalias{2020arXiv201202284C} catalog. For a discussion of the significance of populations in the Galactic Halo, we refer to \citet{2019ApJ...882..169F}(hereafter F19). Here, we want to remark how the broad range in metallicity is an important feature of our samples, as it ensures broad leverage for accurate calibration of the metallicity slope in our period-$\phi_{31}$-[Fe/H] relation.

\noindent

\subsection{Field \rrls{} Optical Data} \label{ssec:optical-data}

The optical ($V$-band) time series data for this analysis have been extracted from the ASAS-SN survey. The first telescope of ASAS-SN came online in 2013, and telescopes have gradually been added for a total of 24 telescopes scattered across the world as of early 2020. At its current capacity, ASAS-SN can survey the entire sky every night, providing high cadence $V$-band photometry, ideally suited to acquire long term, densely populated light curves of \rrls{}. 

\begin{center}
\begin{figure}[!t]
       \includegraphics[width=\textwidth]{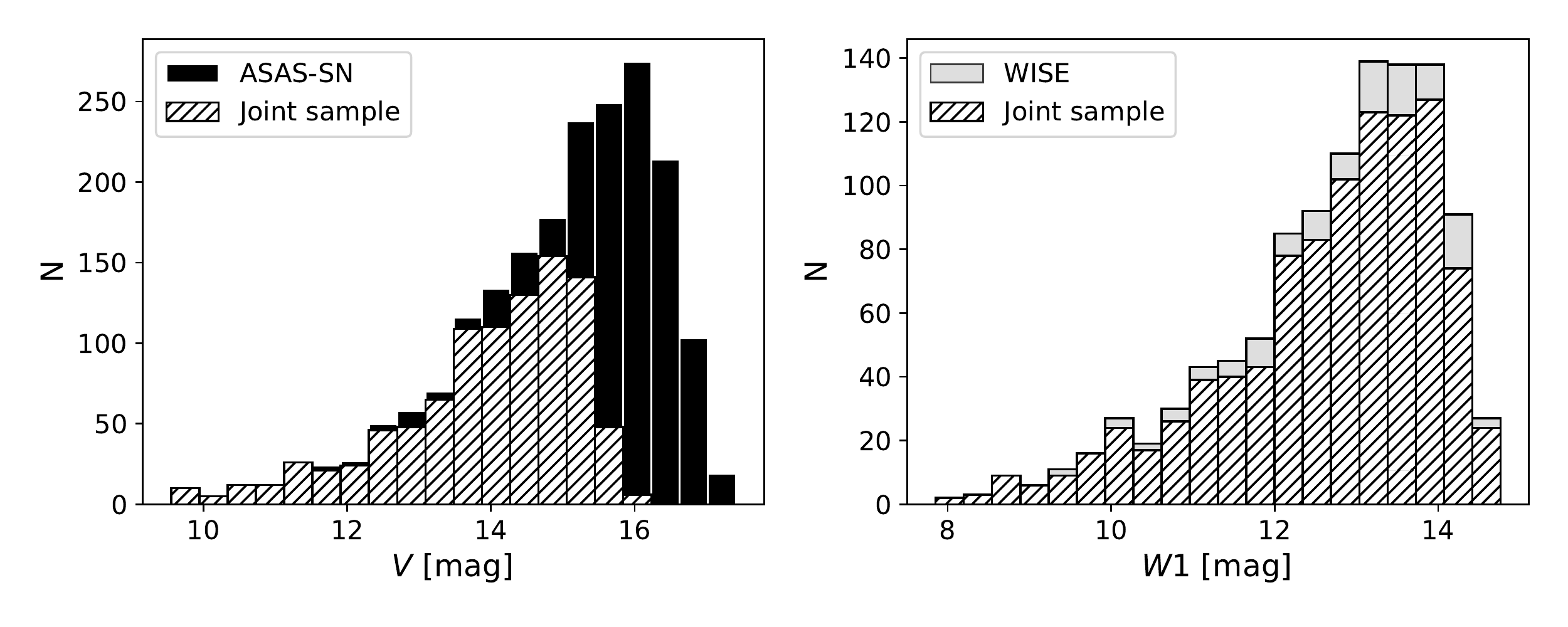}
    \caption{Distribution of the average $V$-band apparent magnitude (left) and the average $W1$-band apparent magnitude (right). Each panel shows both the entire calibration sample (solid fill) and the joint sample (hatched fill, subset of stars having both ASAS-SN and WISE data).}
    \label{fig:mag_distrib}
\end{figure}
\end{center}

Out of the 6079 variables in the HR+$\Delta$S catalog with a match in ASAS-SN, we were able to extract good quality ASAS-SN light curves for 1980 stars using the procedure described in Section~\ref{sec:fourier}. Of these, 967 are in the joint sample, also having good quality light curves in the infrared. The left panel of Figure~\ref{fig:mag_distrib} shows the distribution of average $V$-band apparent magnitudes for both the entire optical dataset (solid black) and the stars in the joint sample (hatched). The cut-off at $V \simeq 17.4$ is due to the limiting magnitude of the ASAS-SN survey. The joint sample is instead truncated at $V \simeq 16$  due to the shallower photometric depth of the WISE survey. For a detailed description of the generation of the light curves from the ASAS-SN time-series, and calculation of average magnitudes for this dataset, we refer to Section~\ref{ssec:GLOESS}.

\subsection{Field \rrl{} Infrared Data} \label{ssec:infrared-data}

We obtained archival infrared photometric time-series from the WISE mission $W1$ and $W2$ bands (3.4 and 4.6~$\micron$ respectively). The primary WISE mission surveyed the entire sky in four infrared bands every six months from January 2010 to the end of the post cryogenic phase in February 2011. The spacecraft was then reactivated in September 2013 in just the $W1$ and $W2$ bands as the NEOWISE mission. Still having sensitivity similar to the full cryogenic phase, NEOWISE, as of early 2020, has yielded 12 additional full-sky survey epochs (having a minimum of 12 measurements per survey epoch). With a minimum of $\sim$156 individual epochs ($\sim$12 from WISE and 144 from NEOWISE) for each position in the sky, collected over a baseline of almost ten years, the combined WISE and NEOWISE photometry provide the most comprehensive full-sky catalog in the infrared. From hereafter in the paper, the combination of photometry from both the primary WISE and ongoing NEOWISE mission will be referred for brevity as WISE.

Out of 6264 sources in the HR+$\Delta$S with a WISE catalog match, we were able to generate 1083 good quality light curves in at least one of the two adopted WISE bands (1083 and 707 stars in the $W1$ and $W2$ bands, respectively). Figure~\ref{fig:mag_distrib} (right panel) shows the WISE distribution of the average $W1$ apparent magnitude, for both the entire infrared dataset (solid grey) and those stars in the joint sample (hatched). Although the WISE sensitivity limit is $\sim 16$~mag in the two bands of interest, the figure shows a sharp drop at $W1 \simeq 14.7$; fainter \rrls{}, while detected in the WISE and NEOWISE catalogs, tend to have noisy light curves that are then rejected by the quality control procedures described in Section~\ref{sec:fourier}. The histogram for the $W2$-band is similar, although the cut-off in the magnitude distribution happens at $W2 \simeq 14.1$ due to the lower sensitivity in this band, resulting in noisier light curves. The magnitude range and other characteristics in the joint sample subset of WISE bands are nearly the same as the complete infrared dataset due to the large overlap with available ASAS-SN data.


\subsection{Globular Cluster Dataset} \label{ssec:cluster-data}
Lastly, we select a dataset separate from those used in deriving our period-$\phi_{31}$-[Fe/H] relations to use as an independent check of this work (see Section~\ref{ssec:clusters}).
This dataset is comprised of eight globular clusters (GCs) homogeneously spread between [Fe/H] = $-1.1$ and $-2.3$~dex with a sizable number of RRab stars. These clusters are listed in Table~\ref{tab:globulars}. The GC dataset comes mainly from the homogeneous photometric database of P. B. Stetson\footnote{\url{https://www.canfar.net/storage/list/STETSON/homogeneous/Latest_photometry_for_targets_with_at_least_BVI} For additional information on the PBS database, readers may also contact Peter B. Stetson directly at \href{mailto:peter.stetson@nrc-cnrc.gc.ca}{peter.stetson@nrc-cnrc.gc.ca}} (hereafter, PBS), except for NGC~3201 data, which comes from \citet{Piersimoni2002}. The data in the PBS database was collected from ground-based telescopes using archival data from 1984 to present. The main telescopes and cameras which contributed the most to the $V$-band data used in this work include the following: AAO LCOGT 1m (CCD), CTIO 0.9m (Tek2K), CTIO 1m (Y4KCam), CTIO 1.3m (ANDICAM), CTIO LCOGT 1m (CCD), Cerro Pach\'on SOAR 4.1m (SOI), La Silla NTT 3.6m (EMMI TK2048EB), La Silla ESO/MPI 2.2m  (WFI), LCO Warsaw 1.3m (8k-MOSAIC), Maunakea CFHT 3.6m (CFH12K), ORM La Palma JKT 1m (EEV7), ORM La Palma INT 2.5m (WFC), SAAO LCOGT 1m (CCD). For further information about the summary of observing runs for each cluster, the bands observed, and the imagers and telescopes used, we direct the reader to the PBS database (previously mentioned).

\begin{table}[!t]
\caption{Globular Clusters}
\label{tab:globulars}
\begin{center}
\begin{tabular}{lccc}
\tableline
Clusters & [Fe/H]$_{C09}$ & RRab stars\tablenotemark{a} & \colhead{Epochs} \\ 
\tableline
NGC 7078 (M15) & $-2.33$ & 64 (64)& 223 \\ 
NGC 4590 (M68) & $-2.27$ & 14 (13) & 41 \\ 
NGC 4833 & $-1.89$ & 11 (11) & 72 \\ 
NGC 5286 & $-1.70$ & 30 (25) & 111 \\ 
NGC 3201 & $-1.51$ & 72 (50) & 80 \\
NGC 5272 (M3) & $-1.50$ & 177 (175) & 167 \\ 
NGC 5904 (M5) & $-1.37$ & 90 (67) & 87 \\ 
NGC 6362 & $-1.07$ & 18 (18) & 80 \\ 
\tableline
\end{tabular}  
\end{center}
\tablenotetext{a}{The number of RRab listed in each GC is according to Clement's catalog \citep{Clement2001}, where the actual number of RRab available with the PBS photometry is in parentheses.}
\end{table}

The spectroscopic metallicities of these GCs (second column in Table~\ref{tab:globulars}) are listed in the scale of \citetalias{Carretta2009}, and the third column of Table~\ref{tab:globulars} shows the total number of fundamental mode \rrls{} in each GC according to Clement's catalog\footnote{\url{http://www.astro.utoronto.ca/~cclement/read.html}} \citep{Clement2001}. In parentheses, we list the number of RRab actually available in the PBS photometry. In addition, the average number of epochs per $V$-band light curve is shown in the fourth column.


\section{Calibration of Period-Fourier-Metallicity relation} \label{sec:fourier}

Before folding the ASAS-SN and WISE time-series into phased light curves, we ensured that they had at least 30 available epochs, had a $S/N>5$, and no quality or contamination flags in their original archive. Most time-series passed this check.

\begin{figure}[!h]
    \centering
    \includegraphics[width=\textwidth]{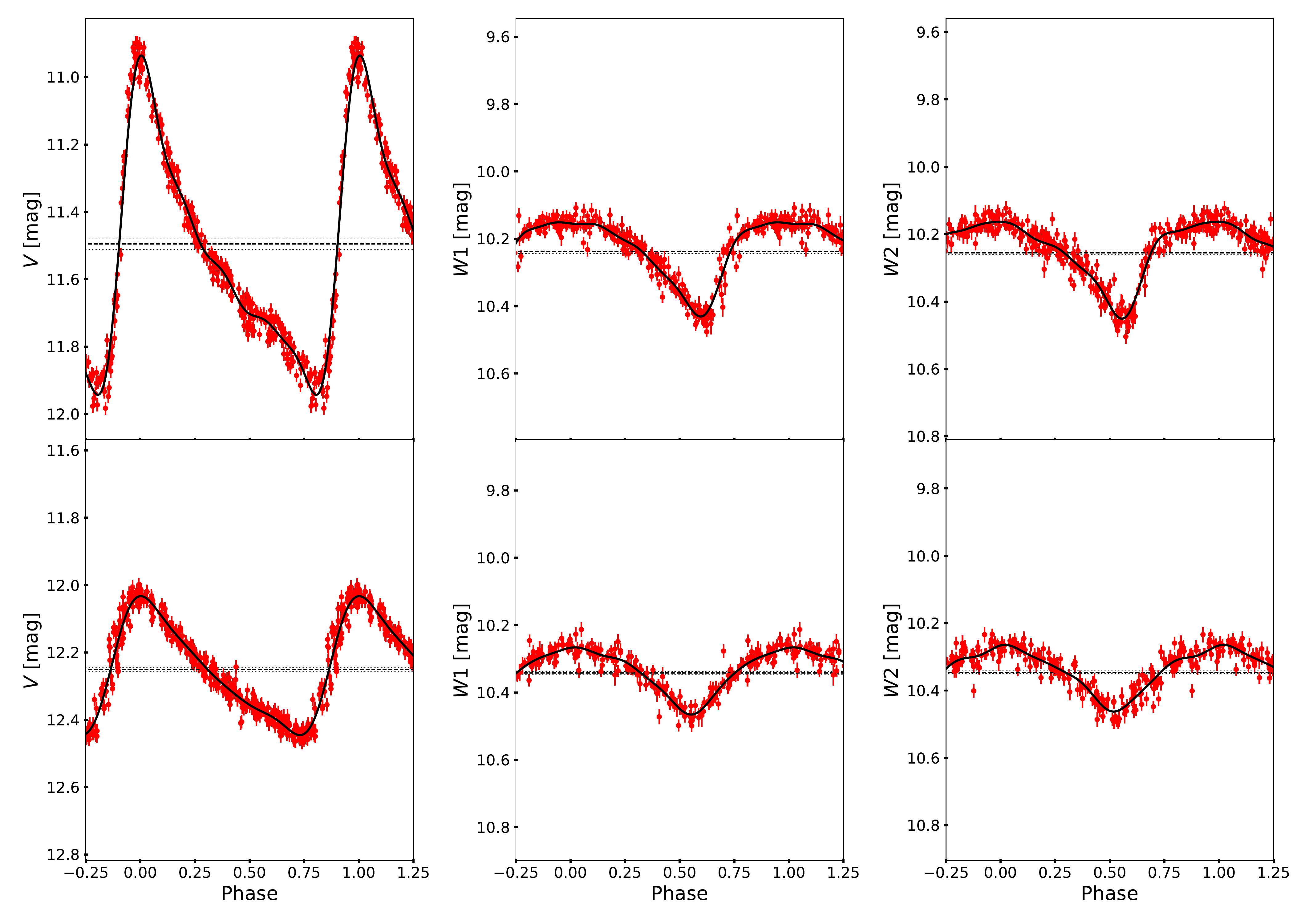}
    \caption{Multiple-band light curves for a typical short and long period star. The $V$ (left), $W1$ (middle), and $W2$ bands (right) are shown for the star DM Cyg (Period = 0.419866 days) in the top row and NSVS 13688631 (Period = 0.737646 days) in the bottom row. The Fourier fit (black solid line) to the GLOESS light curve (see text) is plotted on top of the phased data (red). The average magnitude with its associated error are shown as horizontal black dotted lines. Points automatically rejected by the GLOESS fitting procedure have been removed.}
    \label{fig:LCV}
\end{figure}

Figure~\ref{fig:LCV} shows the $V$, $W1$, and $W2$ light curve typical for two of our stars with good-quality photometry and optimal phase sampling. While the $V$-band has the characteristic saw-tooth shape of RRab variables, with a sharp minimum and maximum connected by a quick rise, the infrared curves are more symmetric and sinusoidal, with a broad maximum, still retaining a sharp minimum. As mentioned above, this is a consequence of optical light curves being more sensitive to changes in the star's effective temperature, while the variations in radius are the determinant factor for the infrared light curves. We take advantage of these wavelength-dependent properties by deriving separate calibrations of the period-$\phi_{31}$-[Fe/H] in the optical and infrared, allowing us to probe different aspects of \rrls{} stellar atmospheric physics.

This section describes the procedure we followed to fit and validate our relations: (a) we first refine the nominal period of each variable found in the literature by taking advantage of the large time-coverage of our ASAS-SN and WISE time series; (b) we then smooth the phased light curves to prepare them for efficient Fourier analysis and perform quality control on the light curves to remove noisy data; (c) we calculate the Fourier decomposition and identify the parameters that are best correlated with metallicity and (d) we finally fit the period-$\phi_{31}$-[Fe/H] relations.

\subsection{Period Determination} \label{ssec:period}

The large temporal baseline of the ASAS-SN and WISE surveys makes our time-series very sensitive to even small errors in the nominal period of the stars. Even inaccuracies of $10^{-6}$~days, over the 10 years span of WISE and its extended mission, result in a $\sim 0.7$\% phase shift when the photometric time series of a 0.55~day period of our typical RRab star is phased, readily detectable in our datasets. To avoid this issue, we have re-derived the periods of all our stars on the basis of their $V$ and $W1$ band\footnote{We did not use the $W2$-band for period determination due to its larger photometric error, often leading to less accurate periods.}  photometry, using the Lomb-Scargle method \citep{1976Ap&SS..39..447L, 1982ApJ...263..835S} within a search window from 0.2 to 1.5 days.

Typical re-derived periods differ from the ones in the literature by $10^{-5}$~days or less. Light curves with period discrepancy larger than $10^{-5}$~days between the $V$ and $W1$ datasets, or with the nominal period from literature, were inspected manually for further quality checks and possible rejection. Approximately 5\% of derived periods fell in this category and required manual inspection. The inability to find a reliable period was also used as a rejection criterion for an individual light curve. Note that this quality assurance process causes the removal of some high amplitude Blazhko stars \citep{Blazhko1907}, due to their naturally larger dispersion in the light curve causing a high ``false alarm'' probability in the period determination returned by the Lomb-Scargle routine. This removal is intended, as the photometric modulation of Blazhko stars can affect the Fourier decomposition of their light curves, complicating the dependence of [Fe/H] from period and $\phi_{31}$ (see e.g. \citealt{2014A&A...562A..90S}).

\subsection{GLOESS Light Curve Smoothing and Quality Control} \label{ssec:GLOESS}

Rather than directly calculate the Fourier decomposition of the phased light curve, we elected to smooth it first to better deal with the uneven sampling of the observing epochs, and primarily to allow the removal of data points with excessive noise. The smoothing was performed using a Gaussian locally-weighted regression smoothing algorithm (GLOESS, \citealt{2004AJ....128.2239P}). This method places the phased light curve on an interpolated grid, to which a second-degree polynomial is locally fit to the photometric data. Fitting weights depend on both the photometric error and the Gaussian distance of the phased photometry from the interpolation point. The procedure was repeated multiple times in order to apply an iterative sigma clipping rejection scheme, designed for noisy data point removal. Only time series with a minimum of 30 valid photometric data points after sigma clipping were retained for processing, with the rest excluded from further consideration. A full description of our implementation of the GLOESS method is available in \citet{2015ApJ...808...11N}.
	
Figure~\ref{fig:LCV} shows example light curves of the phased data (red points) for both a typical short and long period \rrl{} star (DM Cyg and NSVS 13688631, respectively). Data points automatically rejected by the GLOESS fitting procedure are removed. From the GLOESS light curve, we have measured the mean magnitude (calculated as the mean of the smoothed light curve in flux units, converted back to magnitude), the amplitude, and the epoch of maximum (separately in each band). The mean magnitude uncertainty was calculated as the sum in quadrature of the photometric uncertainty and the uncertainty in the fit (see \citealt{2015ApJ...808...11N} for details). The error associated with the amplitude was instead defined as the standard deviation of the data points residuals with respect to the smoothed light curve. Stars with an amplitude equal to less than three times the amplitude error were found to correspond to excessively noisy light curves and were excluded from further analysis. Note that this process also excluded large modulation Blazhko stars, which would appear as noisy light curves with large uncertainty in amplitude.

\subsection{Fourier Decomposition} \label{ssec:Fourier}

Fourier decomposition is a widely adopted tool to quantify the shape of a light curve, as the lower order terms in the expansion are usually sufficient to fully characterize its shape (see e.g. \citealt{1981ApJ...248..291S}). As justified above, for this work we elected to perform Fourier expansion of the evenly spaced smoothed light curves, rather than the individual data points, in the form:

\begin{equation}\label{eq:fourier_expansion}
        m(\Phi)=A_0+\sum_{i=1}^{n} A_{i}\sin[2\pi i  (\Phi+\Phi_{0})+ \phi_{i}]
\end{equation}
    
\noindent
where $m(\Phi)$ is the observed magnitude for either the ASAS-SN or WISE bands, $A_{0}$ is the mean magnitude, $n$ is the order of the expansion, $\Phi$ is the phase from the GLOESS light curve varying from 0 to 1, $\Phi_{0}$ is the phase that corresponds to the time of maximum light $T_{0}$, and the $A_{i}$'s and $\phi_{i}$'s are the $i$-th order Fourier amplitude and phase coefficients, respectively.

We determined the Fourier coefficients with a weighted least-squares fit of Equation~\ref{eq:fourier_expansion} to the smoothed GLOESS light curve and its locally calculated error. The locally calculated error was defined as the local photometric scatter around the smoothed light curve, estimated as the weighted standard deviation of the residual with the data convolved with the GLOESS smoothing kernel. We found that a fifth-order ($n = 5$) Fourier expansion was sufficient to reproduce the shape of the light curve in each band. Examples are shown in Figure~\ref{fig:LCV} for two typical \rrls{}. The Fourier decomposition fit (black solid line) is plotted on top of the actual photometric data in red.

\citet{1981ApJ...248..291S} first showed that certain combinations of Fourier coefficients were directly related to some physical parameters of pulsating stars. These coefficients are typically defined either as linear combinations of Fourier phases:

\begin{equation}\label{eq:phi_def}
     \phi_{ij}=j\cdot \phi_{i}-i \cdot\phi_{j}
\end{equation}

\noindent     
where $\phi_{ij}$ is cyclic in nature and ranges from $0$ to $2\pi$, or as ratios of the Fourier amplitudes:

\begin{equation}\label{eq:Rij}
     R_{ij}=\frac{A_{i}}{A_{j}}
\end{equation}

In Appendix~\ref{sec:Fourier_param_plots}, we discuss the correlation between various combinations of Fourier parameters among themselves, and with period, and how they help discriminate sources with different metallicities. Our analysis confirms the conclusions of early studies such as \citetalias{1996A&A...312..111J}, suggesting that a relation between period and $\phi_{31}$ is a good indicator of metallicity in the $V$-band. We found this to be true also in the infrared. Uncertainties in the calculated $\phi_{31}$ values are several orders of magnitude less than our best uncertainties in metallicity and are therefore deemed negligible throughout the rest of this work.

\begin{figure}[!t]
        \centering
        \includegraphics[scale=0.8]{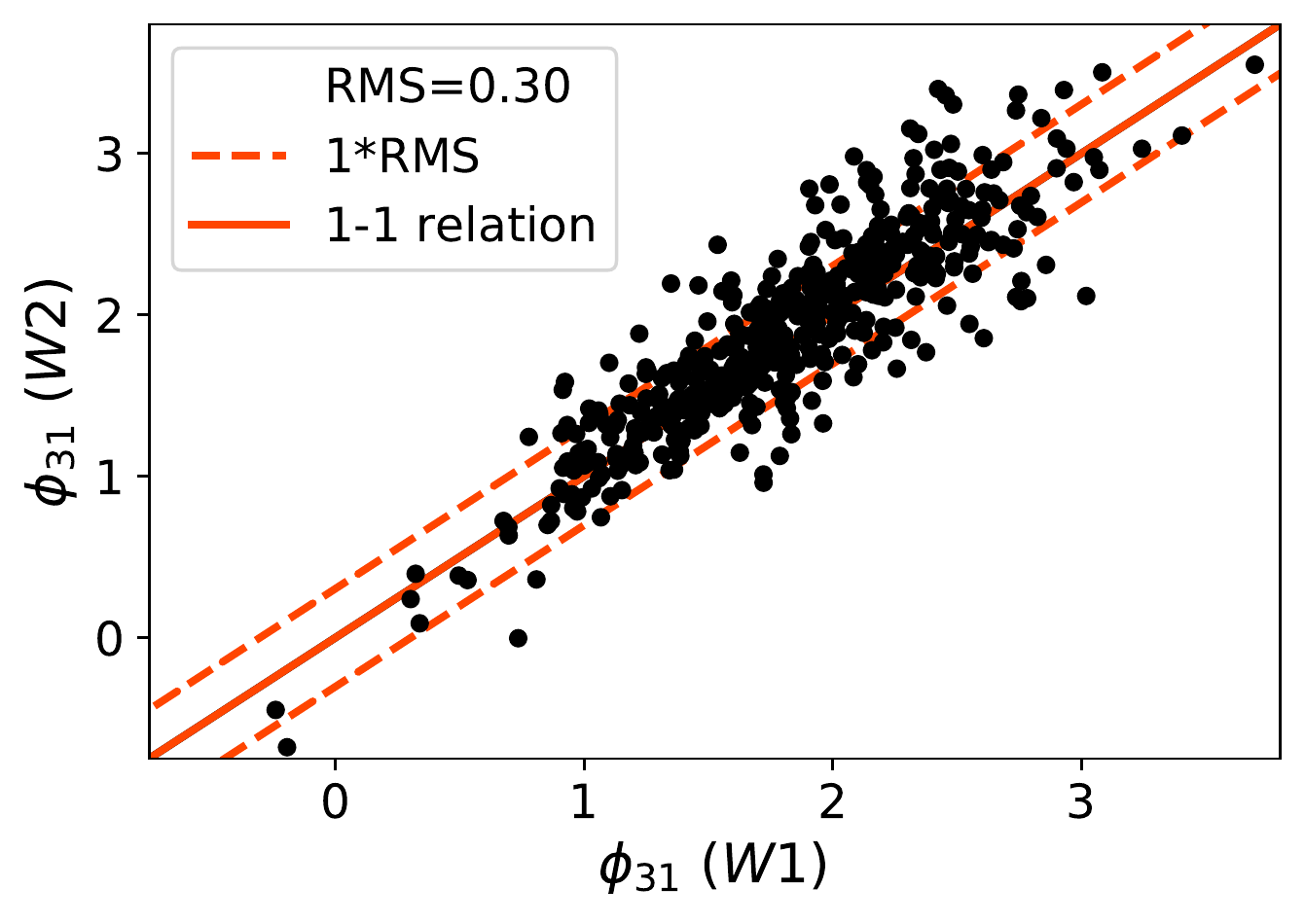}
        \caption{Comparison between $W1$ and $W2$ $\phi_{31}$ parameter. The dashed orange line shows the RMS between the two different $\phi_{31}$ values, or the typical scatter around the ideal 1-1 relation. The datapoints shown correspond to the subset of the WISE dataset with both $W1$ and $W2$ $\phi_{31}$ measurements available.}
        \label{fig:PHI31_COMPARISON}
\end{figure}

Figure~\ref{fig:PHI31_COMPARISON} shows that the $W1$ and $W2$ bands produce indistinguishable values of $\phi_{31}$ due to the light curves in these bands being nearly identical. This is demonstrated quantitatively by Equation~\ref{eq:phi31_interrelation}, which shows that the best-fit slope between $\phi_{31(W1)}$ and $\phi_{31(W2)}$ is within the errors close to unity, with a dispersion of 0.30.

\begin{equation}\label{eq:phi31_interrelation}
    \phi_{31(W2)}=(1.011\pm 0.023)\cdot \phi_{31(W1)}+(0.066\pm 0.043)
\end{equation}

We take advantage of this strong correlation by averaging, whenever possible, the $\phi_{31}$ parameters calculated for the two bands, falling back on $\phi_{31(W1)}$ or $\phi_{31(W2)}$ when only one is available. This essentially doubles the signal of the WISE light curves by combining the data of independent measurements taken at different wavelengths.

\subsection{Period-Fourier-[Fe/H] Fitting} \label{ssec:planefit}

The metallicity of each star in either our ASAS-SN or WISE datasets can be effectively represented by a plane in the period, $\phi_{31}$, and [Fe/H] space, as is demonstrated in Appendix~\ref{sec:PCA} by performing Principal Component Analysis (PCA) on each dataset. To determine the orientation of this plane, we adopted the Orthogonal Distance Regression (ODR) routine part of the SciPy package\footnote{\url{https://docs.scipy.org/doc/scipy/reference/odr.html}}, which utilizes a modified trust-region Levenberg-Marquardt-type algorithm \citep{ODR_Boggs} to estimate the best fitting parameters. We chose ODR because it can be used similarly to PCA with both techniques minimizing the perpendicular distance to the fit with no differentiation between dependent and independent variables. This allows ODR to produce an optimal fit despite the correlation between variables we found in Appendix~\ref{sec:PCA} (especially the well-known correlation between period and metallicity, see e.g. \citetalias{2019ApJ...882..169F}). These correlations, in the case of ordinary least-squares or similar fitting algorithms, would result in a trend of the fitted metallicity residuals with respect to the other variables.

We performed the ODR fit on both our ASAS-SN and WISE calibration sets using the equation:

\begin{equation}\label{eq:plane}
    \textrm{[Fe/H]} = a + b \cdot (P - P_0) + c \cdot(\phi_{31} - \phi_{31_0})
\end{equation}

\noindent
where $P_0$ and $\phi_{31_0}$ are pivot offsets necessary to add an extra element of robustness in the fitting procedure and reduce the fitting parameter uncertainties. We chose the pivot offsets to be near the mean period and $\phi_{31}$ value of each dataset, equal to $P_0 = 0.58$~days for both datasets, and $\phi_{31_0} = 5.25$ and 1.90~radians for ASAS-SN and WISE, respectively. To ensure an accurate fit, two rounds of fitting were performed with an intermediate 4$\sigma$ clipping between the fitted and calibration [Fe/H] (removing $\la 1\%$ of the stars in each calibration dataset).

Our best fit period-$\phi_{31}$-[Fe/H] relation based on the ASAS-SN $V$-band light curves is:

\begin{equation}\label{eq:fourier_ASASSN}
    \textrm{[Fe/H]} = (-1.22 \pm 0.01) + (-7.60 \pm 0.24) \cdot (P - 0.58) + (1.42 \pm 0.05) \cdot (\phi_{31} - 5.25) \  \ ; \ \ \textrm{RMS} = 0.41
\end{equation}

\noindent
where, due to the $2 \pi$ periodic ambiguity in the $\phi_{31}$ coefficient, some $\phi_{31}$ values required adding $2\pi$ to their phase in order to lie closer to the mean $\phi_{31}$ value, as suggested by \citetalias{1996A&A...312..111J}. 

The period-$\phi_{31}$-[Fe/H] relation, based on the average $\phi_{31}$ parameters of the WISE light curves (averaged when possible between the $W1$ and $W2$ bands, as described in Section~\ref{ssec:Fourier}), is instead:

\begin{equation}\label{eq:fourier_NEO}
    \textrm{[Fe/H]} = (-1.47 \pm 0.02) + (-8.33 \pm 0.34) \cdot (P - 0.58) + (0.92 \pm 0.05)\cdot(\phi_{31} - 1.90) \ \ ; \ \ \textrm{RMS} = 0.50
\end{equation}

Parameter uncertainties have been checked with bootstrap re-sampling and are consistent with those via ODR. The errors from bootstrap re-sampling are those listed in the above relations. The Root Mean Square (RMS) of the two relations (0.41 and 0.50~dex for the ASAS-SN and WISE samples, respectively) are similar, showing that indeed accurate photometric metallicities can be obtained from infrared light curves. The RMS values are also comparable to the dispersion that we found with a non-parametric regression scheme, based on the $k$-NN method, of the same data (0.33 and 0.40~dex respectively, see Appendix~\ref{sec:KNN}). This shows that our ODR fits provide an accurate description of the dependence of [Fe/H] from period and $\phi_{31}$, with uncertainty only slightly larger than the data's intrinsic scatter. Table~\ref{tab:phot_table} shows the derived photometric properties for both the $V$-band and mid-IR (WISE) calibration datasets, including the period, $\phi_{31}$ value, and photometric metallicity in each band. Following the light curve fitting, quality control process, and the final plane fit described above, we were left with 1980 variables with good quality ASAS-SN light curves, and 1083 variables with good WISE (in at least one of the two $W1$ and $W2$ bands) light curves.

\begin{table}
\caption{Derived photometric properties\label{tab:phot_table}}
\begin{center}
\begin{tabular}{lccccc}
\tableline
\colhead{Gaia ID} & \colhead{Period\tablenotemark{a}} & \colhead{$\phi_{31}$ ($V$)} & \colhead{[Fe/H]$_{V}$} & \colhead{$\phi_{31}$ ($W$)} & \colhead{[Fe/H]$_{W}$}\\
\colhead{(DR3)} & \colhead{(days)} & \colhead{(radians)} & \colhead{(dex)} & \colhead{(radians)} & \colhead{(dex)}\\
\tableline
\tableline
507222753405440 & 0.6088826 & & & 1.65842 & $-1.93$\\
4235220006525184 & 0.4664141 & & & 1.18099 & $-1.18$\\
5355313117668352 & 0.4945746 & 4.90759 & $-1.06$ & & \\
14233869512030080 & 0.6670034 & 4.96473 & $-2.29$ & 1.62441 & $-2.44$\\
15489408711727488 & 0.6511669 & 5.19695 & $-1.84$ & 1.80280 & $-2.15$\\
15891245851805568 & 0.5738989 & 5.04774 & $-1.46$ & 1.51379 & $-1.77$\\
18268974106572416 & 0.5541744 & 4.75948 & $-1.72$ & 1.26200 & $-1.84$\\
19606736160298112 & 0.5465583 & 5.01736 & $-1.30$ & & \\
20161096179157248 & 0.6042955 & & & 2.31260 & $-1.29$\\
20357148550791168 & 0.4853687 & 4.31030 & $-1.84$ & & \\
\tableline
\end{tabular}  
\end{center}
\tablenotetext{a}{When both $V$-band and mid-IR (WISE) data is present, the period included was calculated from ASAS-SN ($V$-band) data as the period is usually more accurate due to the higher amplitude and steeper light curve.}
\tablecomments{Table~\ref{tab:phot_table} is published in its entirety in the machine-readable format. A portion is shown here for guidance regarding its form and content.}
\end{table}

\begin{figure}[h!]
    \centering
    \includegraphics[width=\textwidth]{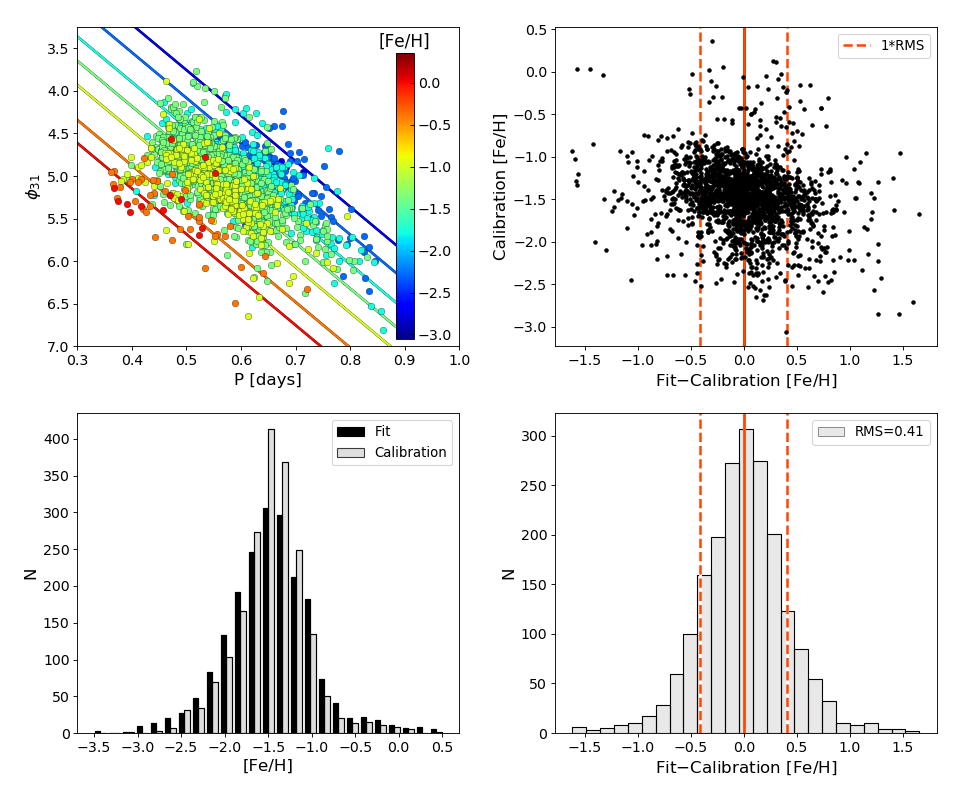}
    \caption{ASAS-SN $V$-band period-$\phi_{31}$-[Fe/H] fit. \emph{Top Left}: period versus $\phi_{31}$ plane with stars binned and color-coded based on their spectroscopic metallicity. Solid lines represent the best fit relation calculated for the average metallicity in each bin, from metal-rich in the bottom left (red) to metal-poor in the upper right (blue). \emph{Bottom Left}: comparison of the spectroscopic calibration [Fe/H] with the best fit photometric metallicity. \emph{Top Right}: spectroscopic [Fe/H] plotted as a function of the residuals of the photometric and spectroscopic metallicity. The dashed vertical lines represent the RMS error between the photometric and spectroscopic metallicity. \emph{Bottom Right}: histogram of the difference between the photometric and spectroscopic [Fe/H]. Again, the vertical lines show the RMS dispersion of the data.}
    \label{fig:ASASSN_ODR}
\end{figure}

\begin{figure}[h!]
    \centering
    \includegraphics[width=\textwidth]{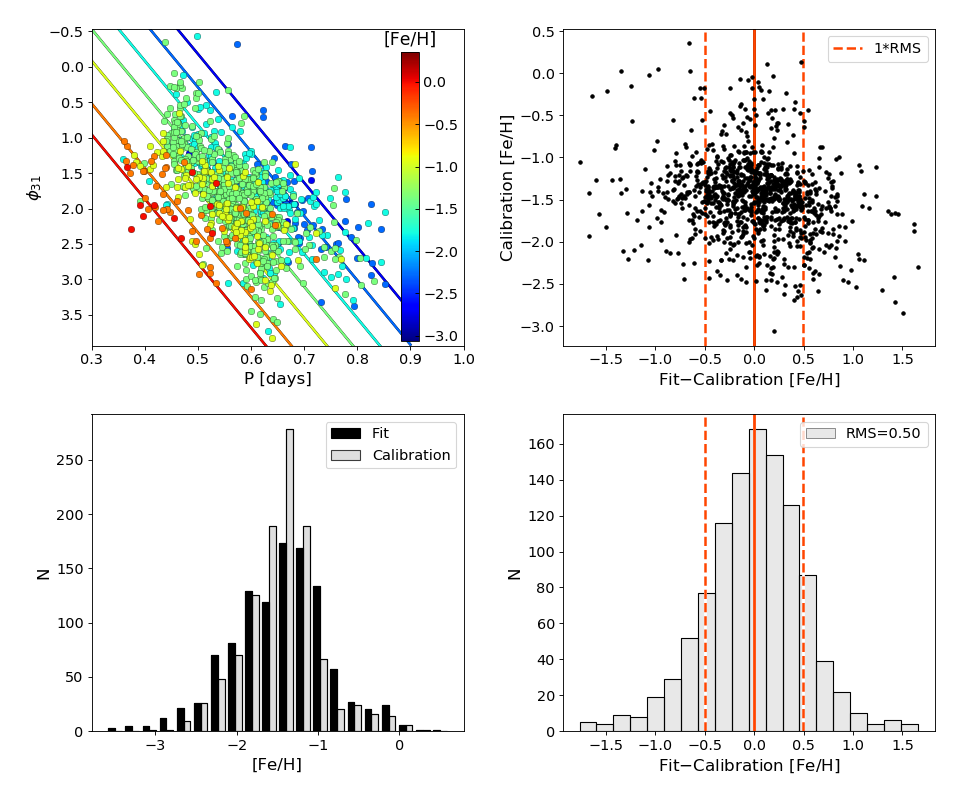}
    \caption{Same as Figure~\ref{fig:ASASSN_ODR}, but for the WISE light curves.}
    \label{fig:WISE_ODR}
\end{figure}

Figures~\ref{fig:ASASSN_ODR} and \ref{fig:WISE_ODR} (for the ASAS-SN and WISE samples, respectively) demonstrate in graphical form the ability of our fits to provide photometric metallicities in agreement with the spectroscopic values in the HR+$\Delta$S calibration sample. The top left panel shows the distribution of the sources in the $\phi_{31}$ vs. period plane, superimposed with the relations in Equation~\ref{eq:fourier_ASASSN} and \ref{eq:fourier_NEO} calculated for fixed values of [Fe/H]. Both data and fit lines are color-binned by metallicity. Note the good agreement between the distribution of the spectroscopic metallicities of individual sources with the locus corresponding to the same metallicity bin defined by the spectroscopic relations. The histogram in the bottom left panel shows how the distribution of the photometric [Fe/H] from our fits closely reproduces the metallicity distribution, from spectroscopy, of the calibration sample. The two plots on the right instead show the distribution of the residuals between the photometric and spectroscopic metallicities: the top panel confirms that our choice of fitting the data with an ODR method indeed avoids residual trends, while the bottom panel shows a close-to-Gaussian distribution of the residuals.

\section{Discussion} \label{sec:discussion}


\subsection{Optical vs. Infrared Relations} \label{ssec:VvsIR}

\begin{figure}[!t]
    \centering
    \includegraphics[width=1.0\textwidth]{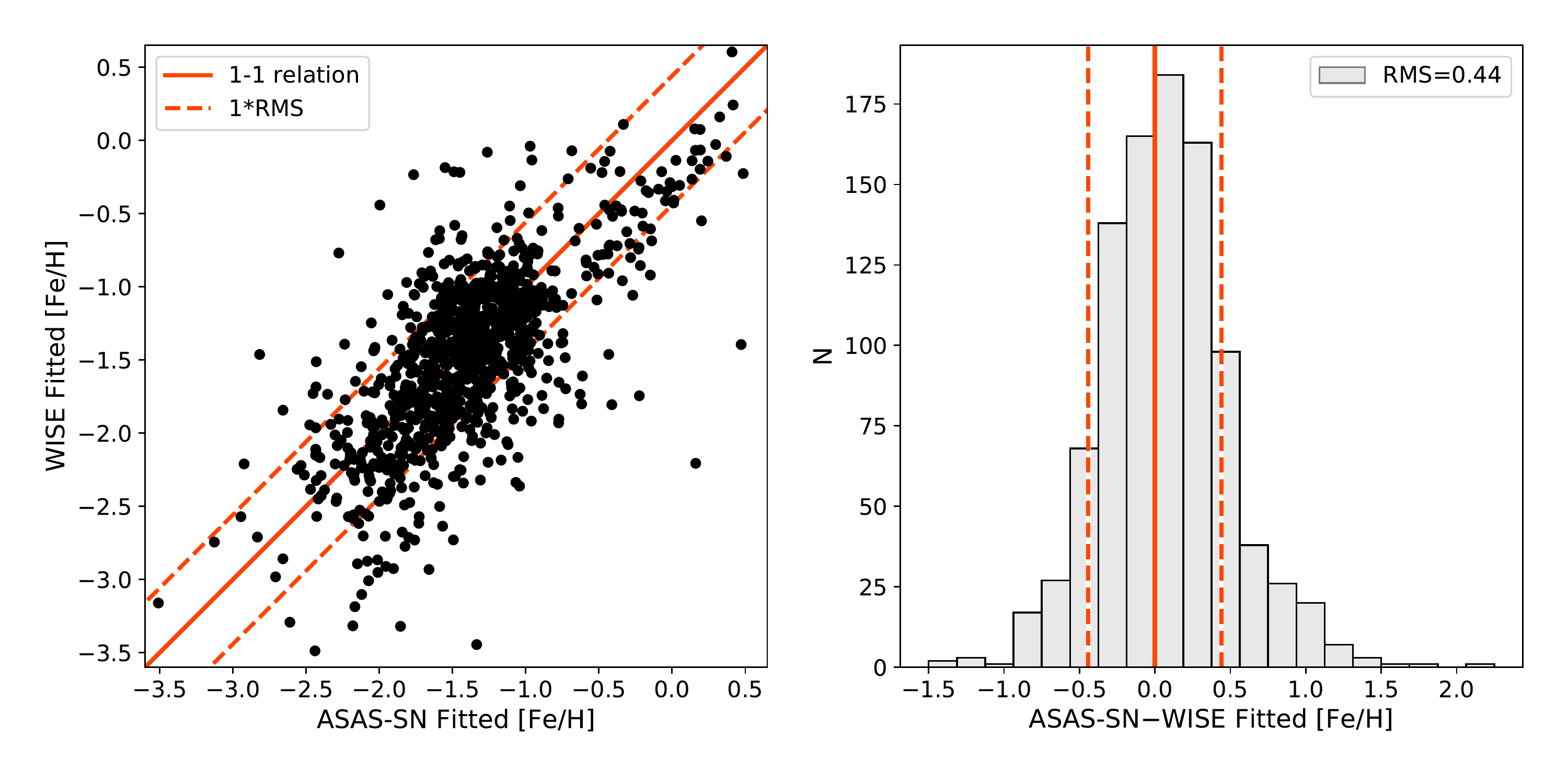}
    \caption{Comparison of the photometric [Fe/H] derived from ASAS-SN and WISE samples. The left panel directly compares the photometric metallicities derived for each star in the joint sample, while the right panel shows a histogram of the differences between these photometric metallicities. The dashed lines show the RMS between the two datasets [Fe/H].}
    \label{fig:WISE_vs_ASASSN_FEH}
\end{figure}

The analysis in Section~\ref{ssec:planefit} shows that optical and infrared period-$\phi_{31}$-[Fe/H] relations provide photometric metallicities of comparable accuracy. However, we still need to test if Equation~\ref{eq:fourier_ASASSN} and \ref{eq:fourier_NEO} provide consistent values of [Fe/H] for individual stars. Figure~\ref{fig:WISE_vs_ASASSN_FEH} shows that this is indeed the case: we find an excellent agreement between the photometric metallicities derived at the two wavelengths ranges for the joint sample on a per-star basis. The dispersion between the two datasets, of 0.44~dex, is comparable with the RMS of the individual fits as well as with the dispersion of the residuals from the $k$-NN method in Appendix~\ref{sec:KNN}.

\begin{figure}[!h]
    \centering
    \includegraphics[width=\textwidth]{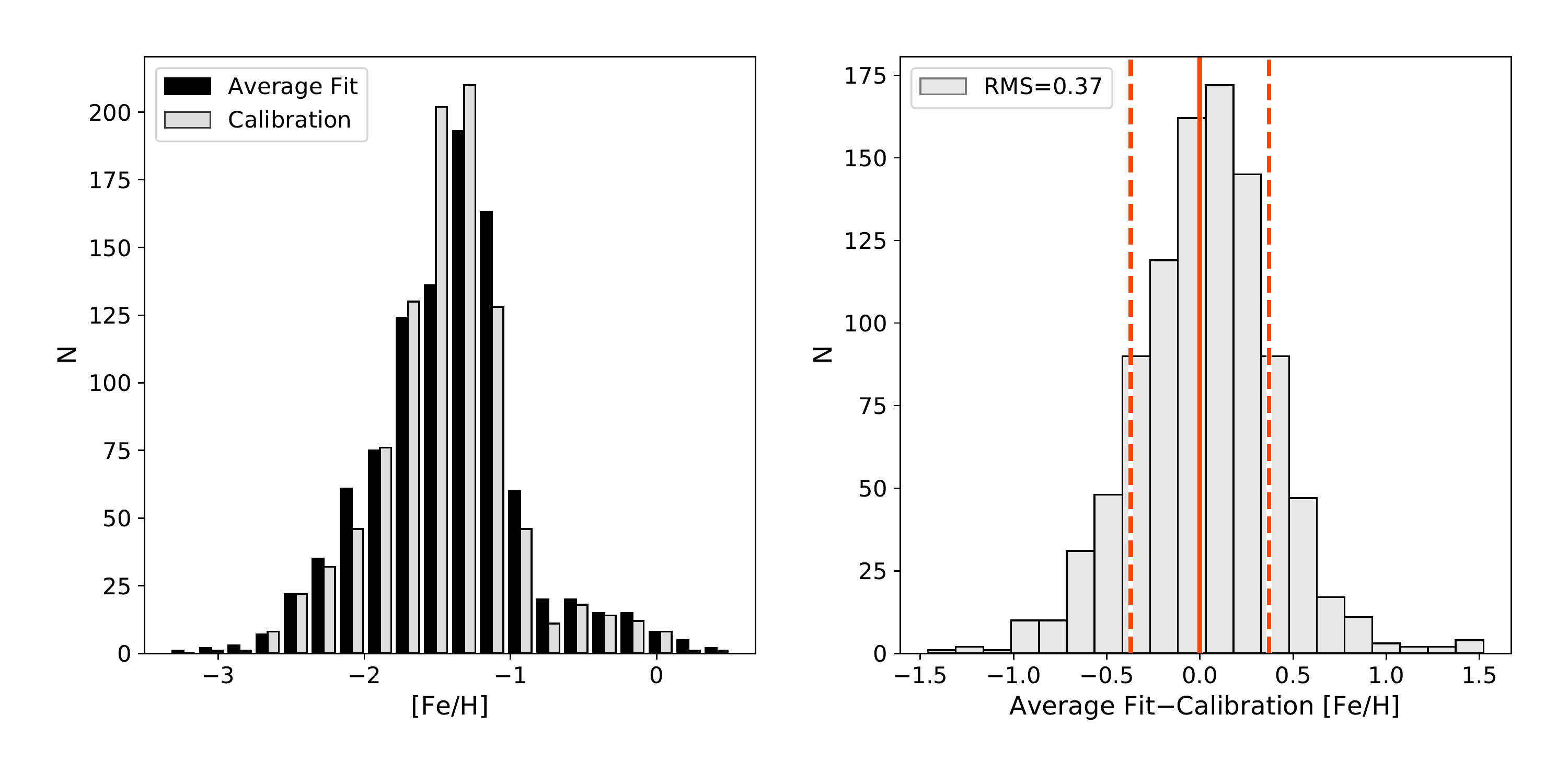}
    \caption{\emph{Left}: Comparison of the distribution of the spectroscopic calibration metallicity (grey histogram) and the fitted photometric metallicity averaged from both the ASAS-SN and WISE $P-\phi_{31}-[Fe/H]$ fit (black histogram). \emph{Right}: Histogram of the difference between the average photometric and spectroscopic metallicities. The vertical orange lines represent the RMS dispersion between the two metallicity values.}
    \label{fig:Averaged_Fit}
\end{figure}

This tight correlation between optical and infrared photometric metallicities, with no apparent trends, led us to consider whether averaging the two photometric metallicities would yield a value significantly closer to the spectroscopic metallicity from the HR+$\Delta$S sample. Furthermore, besides the obvious advantage of increasing the statistics of the photometric measurements, by combining optical and infrared data we can probe the effects that different aspects of stellar atmospheres have on the \rrl{} light curves acquired in these two separate wavelength ranges (the optical light curves are dominated by temperature, while infrared emission follows more closely radius variations).

Based on these arguments, we average together the photometric metallicities derived from the optical and infrared relationships for each star in the joint sample. Figure~\ref{fig:Averaged_Fit} shows that the average photometric metallicity (black histogram in left panel) closely mirrors the spectroscopic metallicity values from the HR+$\Delta$S calibration sample (grey histogram) for the entire metallicity range. The right panel shows that the residuals between the average photometric [Fe/H] and the spectroscopic metallicities fall fairly symmetrically around zero. 

The RMS dispersion between the two sets of values is $\sim$0.37~dex: smaller than the individual error in the optical ($\sim$0.41~dex) or infrared ($\sim$0.50~dex) sample alone. This RMS value is near the dispersion we found with the k-NN method in Appendix C (0.33 and 0.40 dex for the ASAS-SN and WISE datasets, respectively), which shows that combining these datasets allows one to approach the local scatter in the individual datasets.


\subsection{Comparison with Globular Clusters Metallicity} \label{ssec:clusters}

In order to test the $V$-band relation obtained in Section~\ref{ssec:Fourier} on an independent sample, we selected a list of eight Galactic GCs homogeneously spread between [Fe/H] = $-1.1$ and $-2.3$~dex. The selected GCs are described in Section~\ref{ssec:cluster-data} and listed in Table~\ref{tab:globulars}. For each cluster, we chose the RRab stars with the best sampled light curves in the $V$-band, avoiding those RRab stars that suffer the Blazhko effect because of the modulation of the amplitude and the shape of their light curves. A Fourier decomposition was performed on each light curve to obtain their $\phi_{31}$ parameter. The period-$\phi_{31}$-[Fe/H] relation (Equation~\ref{eq:fourier_ASASSN}) was then applied to estimate the metallicity of each cluster star.

\begin{figure}[!t]
    \centering
    \includegraphics[scale=0.4]{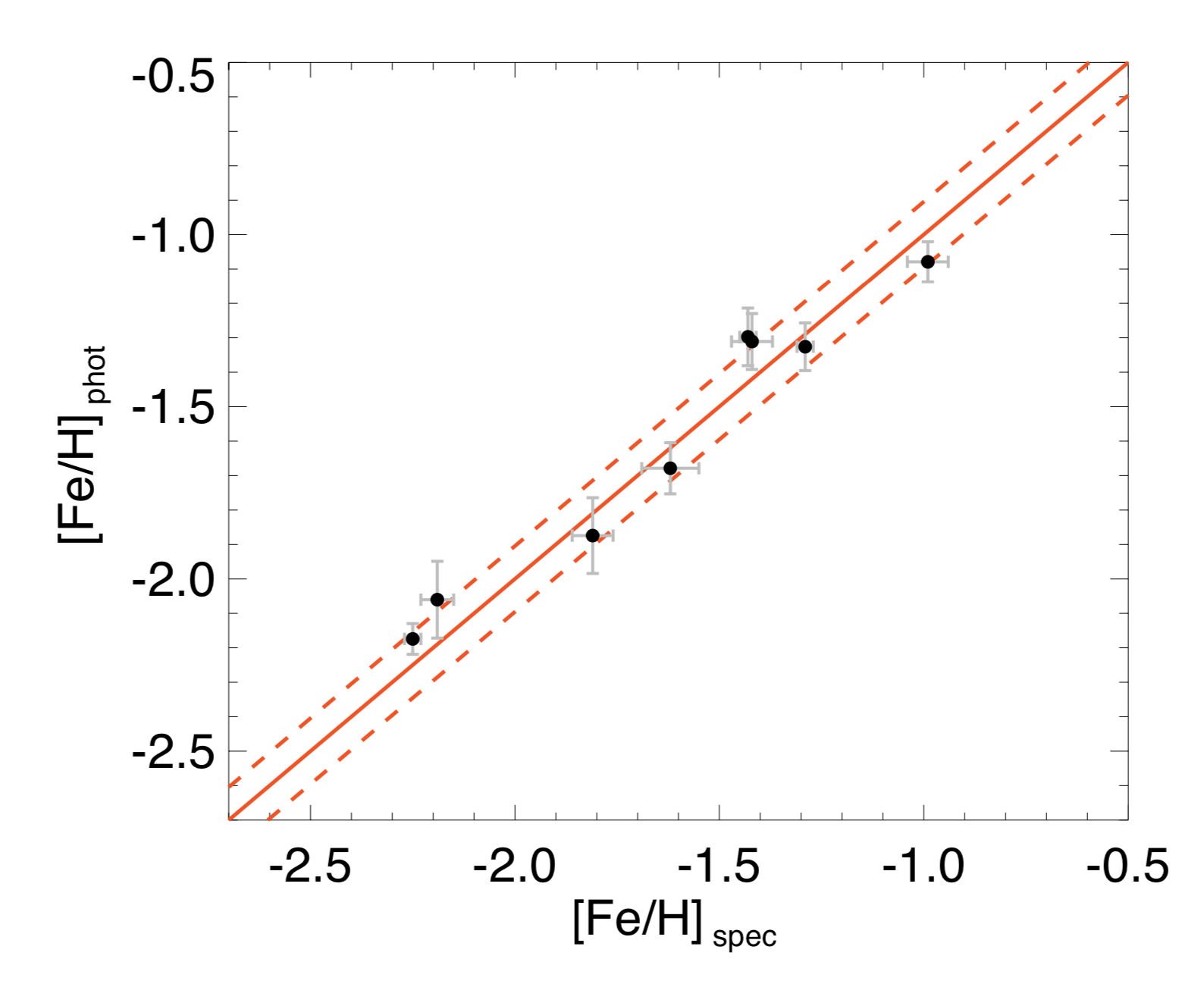}
    \caption{[Fe/H]$_{\rm{spec}}$ versus [Fe/H]$_{\rm{phot}}$ obtained by using Equation~\ref{eq:fourier_ASASSN} on a sample of globular clusters. The solid orange line is the 1-1 relation, while the dashed orange lines show the standard deviation ($\pm$0.09~dex). Error bars correspond to the uncertainties in spectroscopic metallicity and the statistical error in the photometric metallicity.
    }
    \label{fig:feh_comparison_GCs}
\end{figure}

Figure~\ref{fig:feh_comparison_GCs} shows the spectroscopic [Fe/H] versus the mean photometric [Fe/H] values calculated for each GC with our relation.
For consistency with the field \rrls{} described in Section~\ref{ssec:sample}, the spectroscopic metallicities of these GCs (second column in Table~\ref{tab:globulars}) from \citetalias{Carretta2009} were converted into the scale of this paper with the addition of an offset of $0.08$~dex, as noted in \citetalias{2020arXiv201202284C} and described in Section~\ref{ssec:sample}. The error bars correspond to the spectroscopic and photometric metallicity uncertainties. The former comes from \citet{Carretta2009}, while the latter was assessed as the standard error of the mean for star photometric metallicities in each cluster. The good performance obtained using the period-$\phi_{31}$-[Fe/H] relation derived in this work is clearly noticeable in Figure~\ref{fig:feh_comparison_GCs}. There are no signals of possible systematic effects, and  the predicted [Fe/H] using the period-$\phi_{31}$-[Fe/H] relation obtained in this work is within $\pm$0.09~dex of the spectroscopic [Fe/H]. In fact, this accuracy is similar to the metallicity uncertainties measured from the high-resolution spectroscopy on individual stars.

\subsection{Comparison with other Relations}\label{ssec:validation}

\begin{figure}[!t]
        \centering
        \includegraphics[width=\textwidth]{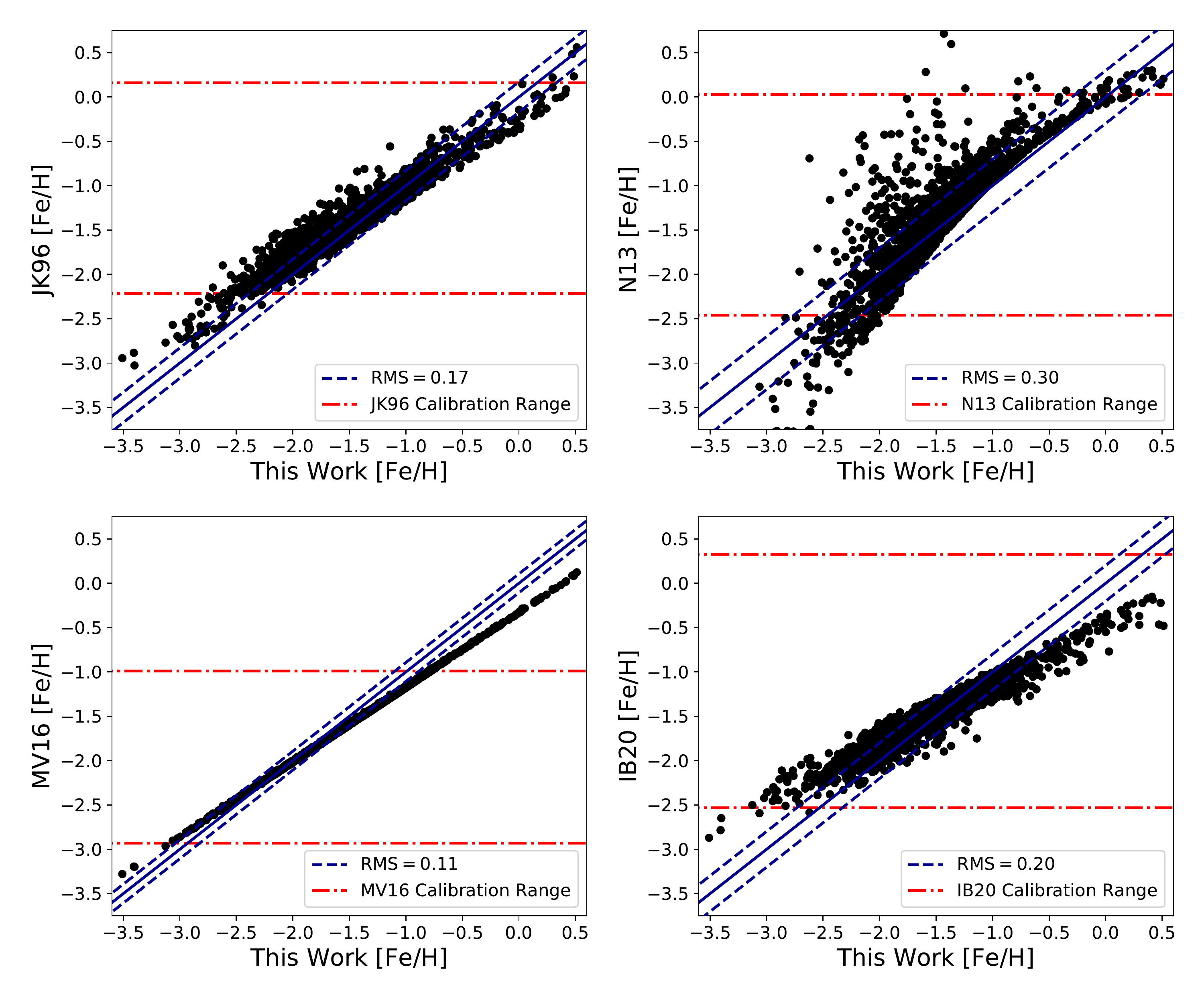}
        \caption{Comparison of the [Fe/H] derived with our $V$-band $P-\phi_{31}-$[Fe/H] relation versus that derived using the relationships of \citetalias{1996A&A...312..111J} (top left), \citetalias{2013ApJ...773..181N} (top right), \citetalias{Martinez-Vazquez2016} (bottom left), and \citetalias{2020arXiv200802280I} (bottom right), applied to the period and $\phi_{31}$ values derived in the ASAS-SN calibration dataset. The range of calibration metallicities used in each work to derive their respective relations is noted in each panel with the two red dash-dotted lines. The blue dashed lines around the 1:1 secant (solid line) represent the RMS dispersion between this works and the literature [Fe/H] abundance for the stars within the red dash-dotted calibration range. Note that in the top right panel, outside the calibration region of  \citetalias{2013ApJ...773..181N}, the quadratic trend deviates significantly from this work for low metallicities (beyond the bounds of this panel).}
        \label{fig:Relations}
\end{figure}

In this section, we compare our optical period-$\phi_{31}$-[Fe/H] relation (Equation~\ref{eq:fourier_ASASSN}) with previous relations found in the literature for similar wavelength ranges. In particular, we focus on the relations found in \citetalias{1996A&A...312..111J}, \citetalias{2013ApJ...773..181N}, \citetalias{Martinez-Vazquez2016}, and \citetalias{2020arXiv200802280I}. Note that a similar comparison for our mid-infrared relation (Equation~\ref{eq:fourier_NEO}) is not possible, since we could not find any previous work studying the relation between metallicity and Fourier parameters at wavelengths longer than the $I$-band \citep{2005AcA....55...59S}. The results are shown in Figure~\ref{fig:Relations}, which plots the [Fe/H] abundances calculated with the periods and $\phi_{31}$ parameters from the ASAS-SN sample, using our relation vs. the relations published in the literature.

The top left panel shows the comparison with the formula found in  \citetalias{1996A&A...312..111J} (their Equation~3). Their relation was derived using a total of 81 field RRab with $V$-band photometry from heterogeneous observations at various sites. Due to lack of phase coverage or excessive noise, a direct Fourier fit in \citetalias{1996A&A...312..111J} was often not available, and individual polynomial fits or small parabolas were fit to light curve segments as needed. Since \citetalias{1996A&A...312..111J} adopted metallicities based on the high-dispersion spectroscopy scale of \citet{1995AcA....45..653J}, for consistency, we first converted the metallicities derived with their relation to the \citetalias{Carretta2009} scale, using the relation from \citet{2011MNRAS.415.1366K}: $\rm{[Fe/H]_{C09}}=1.001 \rm{[Fe/H]_{JK96}}-0.112$.
A scale offset of $0.08$ was then added to transform the \citetalias{Carretta2009} scale to that adopted by this work's HR+$\Delta$S calibration sample (see Section~\ref{ssec:sample}). We found agreement within the calibration range of \citetalias{1996A&A...312..111J} (red horizontal lines), with an RMS value of 0.17~dex, which is smaller than the nominal uncertainty in the calculated [Fe/H] abundances, and only a mildly apparent trend.

\citetalias{2013ApJ...773..181N} introduced a quadratic period-$\phi_{31}$-[Fe/H] relation (their Equation 2), calibrated using stars observed in the Kepler photometric band during the first 970~days of the Kepler mission. Since Kepler, during its nominal mission, surveyed the same region of the sky nearly continuously, each star resulted in having $\sim$350,000 data points spread over $\sim$2,500 pulsation cycles, yielding the best sampled light curves (and Fourier decomposition) currently available for any sample of \rrls{}. However, due to the fixed field of view and shallow depth of Kepler's field, their calibration dataset only had a total of 26 RRab stars with accurate metallicity measurements, nine of which were Blazhko. In order to use the relation of \citetalias{2013ApJ...773..181N}, derived in the Kepler photometric system ($Kp$), with our data, we had to convert the $V$-band $\phi_{31}$ value to the Kepler system by adding the systematic offset derived by~\citet{2011MNRAS.417.1022N}: 

\begin{equation}\label{eq:NEMEC2011}
    \phi_{31}(V)=\phi_{31}(Kp)-(0.151\pm0.026)  
\end{equation}

Furthermore, we converted the [Fe/H] abundances provided by \citetalias{2013ApJ...773..181N} from their adopted metallicity scale \citetalias{Carretta2009} to this work scale, using the previously noted scale offset of 0.08 dex. The results are presented in the top right panel of Figure~\ref{fig:Relations} and show an excellent agreement for higher metallicities in the \citetalias{2013ApJ...773..181N} calibration range ($-1.5 \la \textrm{[Fe/H]} \leq 0.03$~dex). For lower metallicities, however, the two relations rapidly diverge (RMS $\simeq 0.30$~dex if calculated over their entire calibration range), caused by the higher-order term in \citetalias{2013ApJ...773..181N}, and possibly due to the scarcity of calibrators in their samples for low [Fe/H] (they only have one \rrl{} with $\textrm{[Fe/H]} \la -2.0$~dex.

Next, we compare our relationship with the one derived by \citetalias{Martinez-Vazquez2016} (end of Section 2 in their paper), calibrated using a sample of 381 \rrls{} in globular clusters binned by period, with the addition of 8 field \rrls{} chosen to extend the metallicity range of the sample. The \citetalias{Carretta2009} metallicity scale used by \citetalias{Martinez-Vazquez2016} has also been converted to that of this work to allow a comparison with our sample. The result is presented in the bottom-right panel of Figure~\ref{fig:Relations}, showing a general agreement with our fit within the calibration range of \citetalias{Martinez-Vazquez2016}, albeit with a small slope with respect to our relation. Note that we set the lower limit of the calibration range plotted to $\textrm{[Fe/H]} \simeq -0.99$ since there was a discontinuity in the data beyond which only had two variable stars.

Finally, \citetalias{2020arXiv200802280I} introduced a $G$-band period-$\phi_{31}$-[Fe/H] relation (their equation 3), based on Gaia DR2 light curves. The relation was calibrated with a sample of 84 stars with known spectroscopic metallicity. To allow comparison with our metallicities, we had first to convert the $V$-band $\phi_{31}$ value of our dataset to the $G$-band system, using the relation derived by \citealt{2016A&A...595A.133C}:

 \begin{equation}\label{eq:SOSDR1}
    \phi_{31}(G)=(0.104\pm0.020)+(1.000\pm0.008)\phi_{31}(V);~\rm{RMS} = 0.055
\end{equation}

Furthermore, an additional $\pi$ offset had to be subtracted from $\phi_{31}$ to set the coefficient on the same scale as \citetalias{2020arXiv200802280I}. Metallicity abundances from \citetalias{2020arXiv200802280I} were on the scale of \citet[][ZW84]{ZinnWest1984} and were thus converted using the following equation from \citetalias{Carretta2009} with a subsequent scale conversion to that of this work: $\rm{[Fe/H]_{C09}}=1.105\rm{[Fe/H]_{ZW84}}+0.160$.

The results are shown in the bottom right panel of Figure~\ref{fig:Relations} and exhibit a clear trend between the two relations for the entire range of metallicity. It should be noted that although the stars used for the calibration of the  \citetalias{2020arXiv200802280I} relation range from ($-2.53 \leq \textrm{[Fe/H]} \leq 0.33$), the fit lacks a significant number of calibrators at the low and high metallicity ends. In comparison to this work, their relation tends to overestimate the metallicity at the metal-poor end and underestimate the metallicity at the metal-rich end.


\subsection{Comparison with High-Resolution Spectroscopic Metallicities}\label{ssec:SpecValidation}

\begin{figure}[!h]
        \centering
        \includegraphics[width=\textwidth]{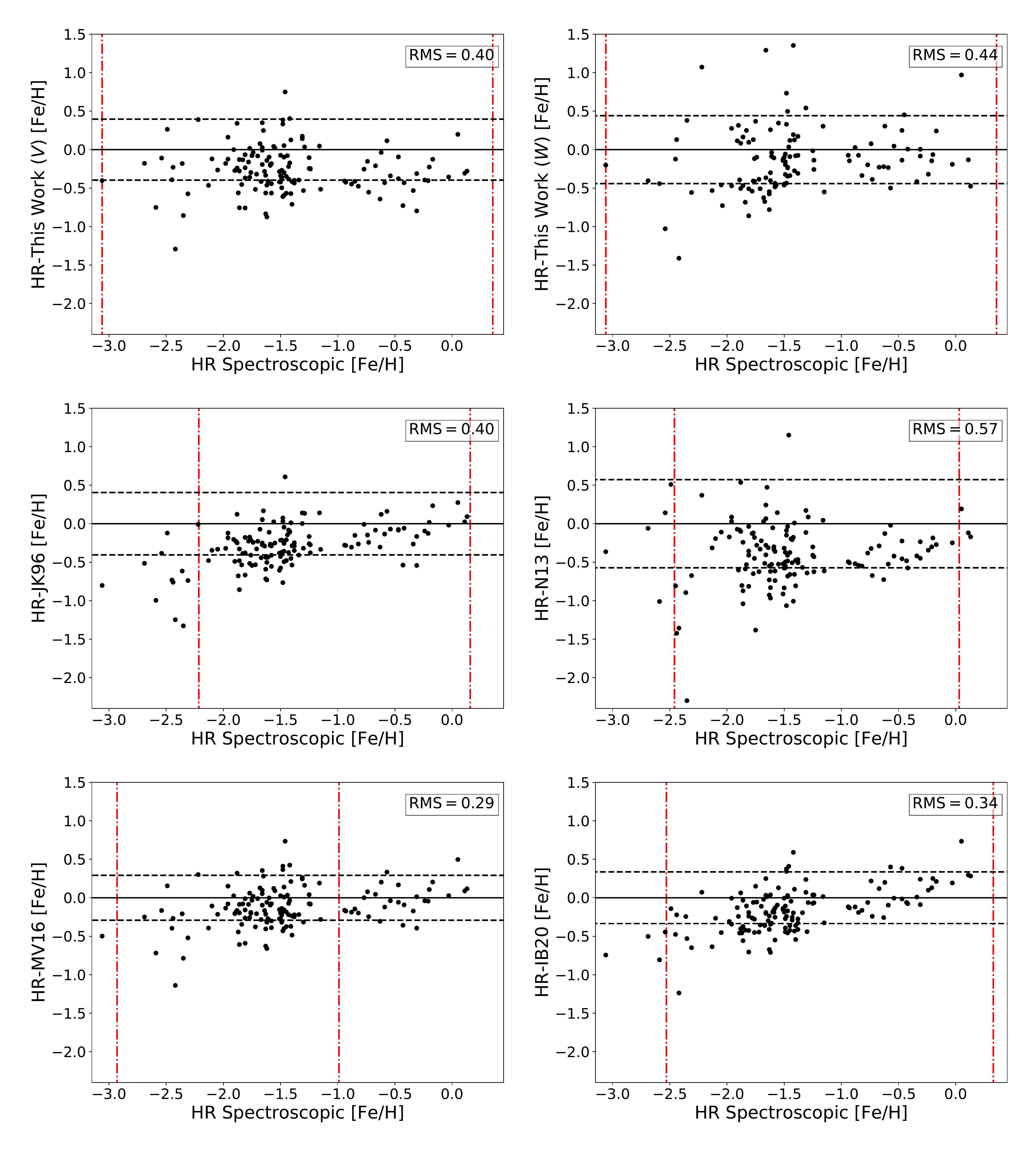}
        \caption{Comparison of the difference between the high-resolution (HR) spectroscopic [Fe/H] and that derived with the photometric relations discussed in Section~\ref{ssec:validation}. The black dashed lines around the zero residual solid line represent the RMS dispersion between the respective work's [Fe/H] abundance and the spectroscopic metallicities from high-resolution data. The range of calibration metallicities used in each work to derive their respective relations is noted in each panel with the two red dash-dotted lines. Note, the mid-infrared comparison (top-right panel) can only be shown for a smaller subset of HR stars for which the WISE light curve is available.}
        \label{fig:SpecRelations}
\end{figure}

To assess the reliability of the period-$\phi_{31}$-metallicity relations, both derived in this paper and from literature, in providing [Fe/H] abundances, we compare their predictions with metallicities measured from high-resolution spectroscopy in the HR+$\Delta$S sample. The results are shown in Figure~\ref{fig:SpecRelations}, plotting the difference between spectroscopic and photometric [Fe/H] abundance for each of the stars in \citetalias{2020arXiv201202284C} with metallicity from a high-resolution spectrum. The RMS scatter of each relation, calculated with respect to zero residuals over the entire spectroscopic metallicity range, is indicated for each photometric relation. 

The figure shows that all relations have similar residuals ($\textrm{RMS} \simeq 0.3$-$0.4$~dex). The scatter in our relations (top row), in particular, are consistent with the expected uncertainties of our photometric metallicities. There is marginal evidence for a systematic shift downwards in panels of Figure~\ref{fig:SpecRelations} that we attribute to the minor differences between the $\Delta$S and HR metallicities as shown in Figure 9 of \citetalias{2020arXiv201202284C}. All relations from literature show a trend of the residuals with high-resolution spectroscopic metallicity due to their tendency to overestimate [Fe/H] abundances at the low-metallicity range and underestimate high-metallicities, with the exception of \citetalias{2013ApJ...773..181N} that shows a significantly larger scatter ($\textrm{RMS} \simeq 0.57$~dex). This apparent trend may result from an inadequate sampling of the metal-rich and metal-poor tails or possibly non-LTE effects in each of the calibration datasets’ metallicities that are not fully taken into account. More investigation is needed in order to clarify the mechanisms behind this trend. This trend is largest in the \citetalias{2020arXiv200802280I}, which underestimates the metallicity of all stars with $\textrm{[Fe/H]} \la -2.0$ and overestimates the metallicity for $\textrm{[Fe/H]} \ga 0.7$~dex, which was already partially noted by \citetalias{2020arXiv200802280I} within their calibration range. 

It is worth noting that the RMS values presented in Figure~\ref{fig:SpecRelations} are generally larger than those quoted inside these respective works of literature. The RMS values quoted in other papers are a measure of their fit's dispersion with respect to their calibration dataset and should not be construed as a measurement of accuracy between two relations. A comparison between works needs to be made on an identical validation dataset (as was done in this section). We attribute the larger dispersion in our relations as denoting a more realistic representation of the intrinsic scatter found in the period-$\phi_{31}$-[Fe/H] relation, which is not apparent in smaller calibration samples, usually covering a smaller metallicity range. Our calibration sample is 76$\times$ larger than \citetalias{2013ApJ...773..181N} (26 RRab) and 24$\times$ larger than \citetalias{1996A&A...312..111J} (81 RRab).  We could decrease the dispersion of this works' relations by artificially cutting the tails of the metallicity distribution (very metal-poor/metal-rich). However, this approach would introduce severe systematic drifts in the metallicity estimates — i.e. the same problem affecting other current calibrations — which is what we have strived to resolve in this work.

\section{Conclusions} \label{sec:conclusions}

In this work, we have calibrated a new relation to provide RRab photometric metallicities based on the $\phi_{31}$ Fourier parameter of optical and, for the first time, infrared light curves. Our relations are based on a set of homogeneous spectroscopic [Fe/H] abundances derived by using the $\Delta S$ method of \citetalias{2020arXiv201202284C} and validated with a sample of [Fe/H] metallicities from high-resolution spectra. The photometric time series of our calibration stars were extracted from the ASAS-SN ($V$ band) and WISE (NEOWISE extension, $W1$ and $W2$ bands) surveys, providing well-sampled light curves that allow for reliable Fourier expansions and accurate determination of the \rrl{} pulsation periods (better than $10^{-6}$~days).

Comparisons with other optical photometric metallicity relations available in the literature show that our relation provides reliable [Fe/H] abundances without noticeable trends over the broadest metallicity range (from $\textrm{[Fe/H]} \la -3$ to solar). We have also shown that there is an intrinsic scatter in the period-$\phi_{31}$-[Fe/H] plane that becomes apparent with large calibration datasets covering a broad metallicity range. Our $V$-band relation (Equation~\ref{eq:fourier_ASASSN}) is consistent within $\simeq 0.40$~dex with [Fe/H] abundances from high-resolution spectroscopy over the entire metallicity range. This relation allows for a quick determination of reliable metallicities for large photometric samples of \rrls{} that will be observed in upcoming optical wide-area time-domain surveys. It offers a good compromise between efficiency and accuracy, with uncertainties only 2 or 3 times larger than metallicities from labor-intensive line-fitting high-resolution spectroscopy. Our relation can also be applied to existing datasets of well-sampled optical light curves obtained by Kepler and Gaia, by converting the $\phi_{31}$ parameters from their respective $K_P$ and $G$ bands, using transformations available in the literature. Finally, we show that our $V$-band relation, applied to ensembles of \rrls{} in Galactic globular clusters, provide estimates of the clusters' metallicity with accuracy comparable to high-resolution spectroscopy measurements (within $\pm 0.09$~dex).

In addition, we have obtained for the first time a mid-infrared period-$\phi_{31}$-[Fe/H] relation using WISE W1 and W2 bands (Equation~\ref{eq:fourier_NEO}). Despite having a smaller number of RRLs to derive this relation, the RMS obtained was similar to that we obtained in the optical. While our infrared relations are slightly less accurate ($\textrm{RMS} \simeq 0.50$~dex), they can still be used to obtain statistically representative metallicities for large ensembles of \rrls{}. This will be crucial with the advent of sensitive telescopes in the mid-infrared (such as JWST), which will allow observations of extragalactic \rrls{} across the Local Group of galaxies, for which spectral observations will not be feasible. Further new large optical (WEAVE\footnote{\url{https://ingconfluence.ing.iac.es:8444/confluence//display/WEAV/The+WEAVE+Project}}, 4MOST\footnote{\url{https://www.4most.eu/cms/facility/overview/}}) and near-infrared (MOONS\footnote{\url{https://www.eso.org/sci/facilities/develop/instruments/MOONS.html}} at VLT) spectroscopic surveys will provide the unique opportunity to improve the current calibration of the period-Fourier-metallicity relations using both medium and high-resolution spectra. However, by combining optical and infrared photometric metallicities, we have shown it is possible to further improve the measurements' reliability.

Having a larger sample size of \rrl{}s with known metallicity will enable a more in-depth study of the metallicity distributions of \rrl{}s (tracers of older stellar populations) in our local neighborhood.


\acknowledgments
This publication makes use of data products from WISE, which is a joint project of the University of California, Los Angeles, and the Jet Propulsion Laboratory (JPL)/California Institute of Technology (Caltech), funded by the National Aeronautics and Space Administration (NASA), and from NEOWISE, which is a JPL/Caltech project funded by NASA’s Planetary Science Division. 

This publication also makes use of data products from the ASAS-SN project, which has their telescopes hosted by Las Cumbres Observatory. ASAS-SN is supported by the Gordon and Betty Moore Foundation through grant GBMF5490 and the NSF by grants AST-151592 and AST-1908570. Development of ASAS-SN has been supported by Peking University, Mt. Cuba Astronomical Foundation, Ohio State University Center for Cosmology and AstroParticle Physics, the Chinese Academy of Sciences South America Center for Astronomy (CASSACA), the Villum Foundation, and George Skestos.

This work has made use of data from the European Space Agency (ESA) mission
{\it Gaia} (\url{https://www.cosmos.esa.int/gaia}), processed by the {\it Gaia}
Data Processing and Analysis Consortium (DPAC,
\url{https://www.cosmos.esa.int/web/gaia/dpac/consortium}). Funding for the DPAC
has been provided by national institutions, in particular the institutions
participating in the {\it Gaia} Multilateral Agreement.

M. Marengo and J. P. Mullen are supported by the National Science Foundation under Grant No. AST-1714534. M. Monelli has been supported by the Spanish Ministry of Economy and Competitiveness (MINECO) under the grant AYA2017-89076-P.

%

\vspace{5mm}
\facilities{WISE, ASAS-SN, \textit{Gaia}}


\software{Astropy \citep{2013A&A...558A..33A}, SciPy \citep{2020SciPy-NMeth}, scikit-learn \citep{scikit-learn}}







\bibliographystyle{aasjournal}

\appendix

\section{Fourier Parameters}\label{sec:Fourier_param_plots}

\begin{figure}[!h]
    \centering
    \includegraphics[width=\textwidth]{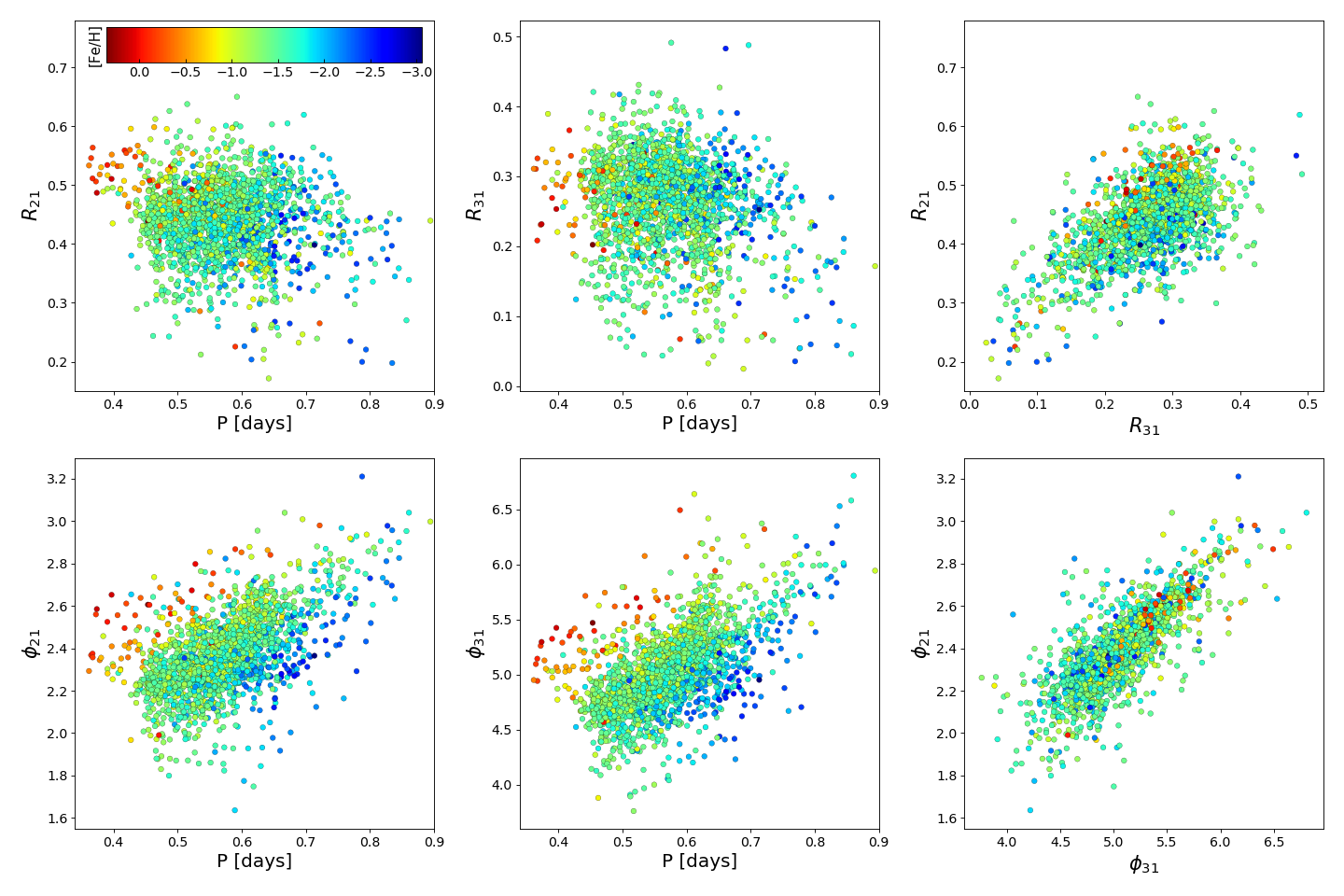}
    \caption{ASAS-SN $V$-band Fourier amplitude parameter ratios (top row) and linear phase parameter combinations (bottom row). \rrls~are color-coded based on their spectroscopic calibration metallicities. In the center and left panels, [Fe/H] generally goes from metal-rich in the left (red) to metal-poor in the right (blue).}
    \label{fig:ASASSN_FOURIER_COEFFS}
\end{figure}

\begin{figure}[!h]
    \centering
    \includegraphics[width=\textwidth]{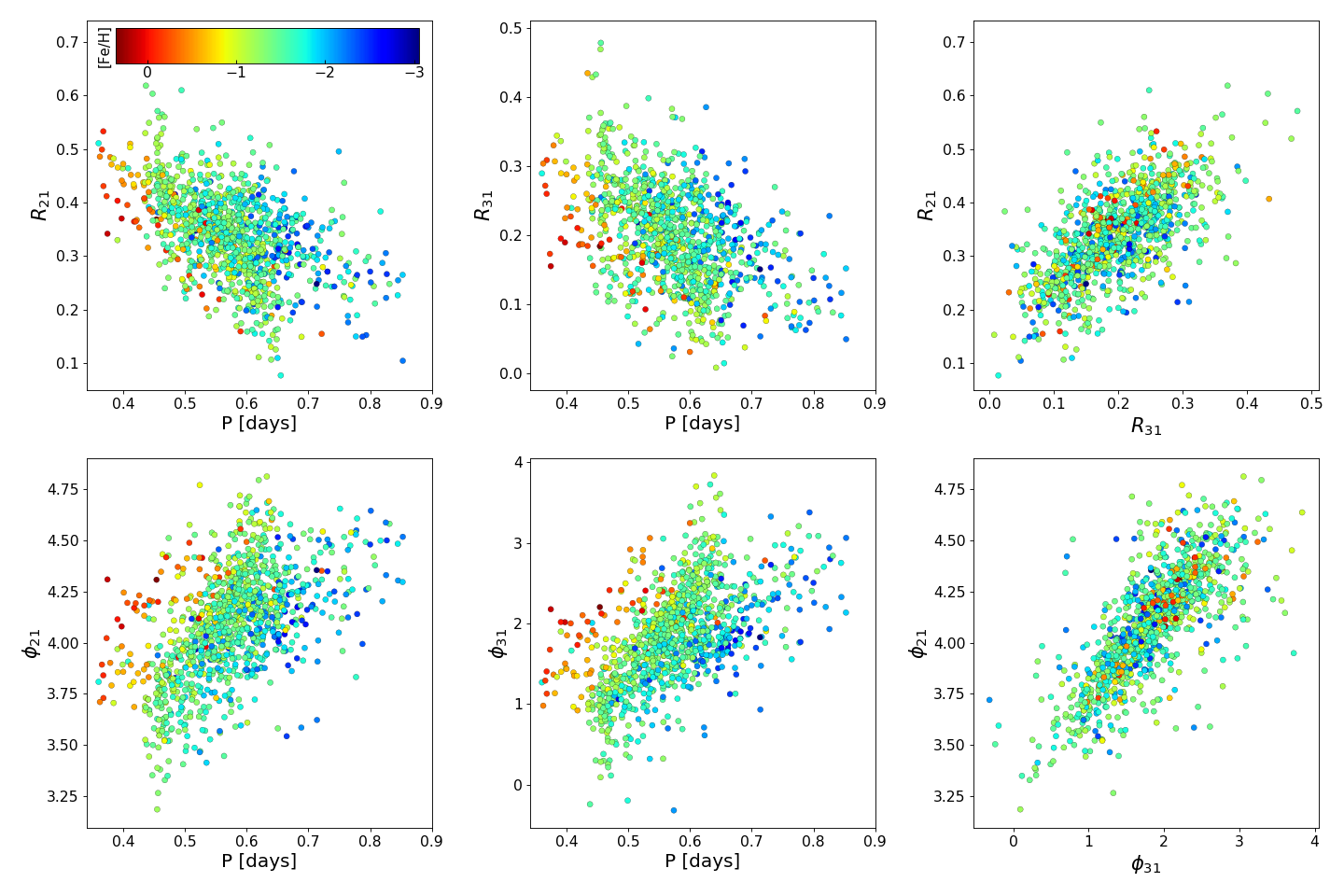}
    \caption{Same as Figure~\ref{fig:ASASSN_FOURIER_COEFFS}, but for our WISE sample.}
    \label{fig:WISE_FOURIER_COEFFS}
\end{figure}

To visualize if the shape of an infrared \rrl{} light curve retains information about the star's metallicity (as is the case in the optical), we have analyzed several low order Fourier parameters, plotted one against the other, or vs. the period. Figures~\ref{fig:ASASSN_FOURIER_COEFFS}  and \ref{fig:WISE_FOURIER_COEFFS} plot individual \rrls, color-coded on their [Fe/H], on the basis of $R_{21}$, $R_{31}$, $\phi_{21}$, $\phi_{31}$, and period for the $V$ and $W1$ bands respectively. In all panels where a Fourier parameter is plotted as a function of period, we can readily see a gradient in the [Fe/H] distribution of the stars.

The stronger separation is apparent in the $\phi_{21}$ and $\phi_{31}$ vs. period plots, as already discussed in \citetalias{1996A&A...312..111J} for $V$-band light curves. We found that the same is true in the WISE $W1$ and $W2$ bands. In accordance with previous literature, and based on the observation that the $\phi_{31}$ parameter provides a tighter relation with period and metallicity at both optical and infrared wavelengths, we have adopted $\phi_{31}$ as the parameter of choice for the analysis presented in Section~\ref{ssec:Fourier}.

\section{Principal Component Analysis} \label{sec:PCA}

Previous works have successfully determined linear relations between period, $\phi_{31}$, and [Fe/H] abundance in RRab stars (except for \citetalias{2013ApJ...773..181N}, which preferred a quadratic form for the relation). In the case of uncorrelated variables, coefficients of such relations can be estimated with a simple least-squares fit of the available data. It is well known, however, that there exists a correlation between period and metallicity in RRab stars (see e.g. Equation~14 in \citetalias{2019ApJ...882..169F}). To explore this correlation and test if the distribution of metallicity can be effectively described with a plane in period and $\phi_{31}$ not just at optical wavelengths, but also in the infrared, we have performed a Principal Component Analysis (PCA) to these three dimensions of our data.

PCA (see e.g. \citealt{DBLP:journals/corr/Shlens14}) is a machine learning algorithm often utilized for reducing the number of variables needed to describe a data set. Given a data set with $p$-different variables, PCA finds a vector (principal component) in this $p$-dimensional space that can explain the most variance. Subsequent components are found orthogonal to the prior components in a direction that explains the next highest amount of variance. The percent variance of each of these components quantifies their relative importance. In other words, PCA gives the ability to locate an $n$-dimensional hyperplane in the $p$-dimensional variable space ($n<p$, with the smallest possible $n$) that characterizes the largest amount of variance in the data.

The first step for PCA analysis is to re-normalize each variable (period, $\phi_{31}$, and [Fe/H] abundance) in both samples (optical and infrared) so that they have zero mean and unitary variance. This avoids common issues in PCA when different variables range in scale. The principal components were then found by utilizing the PCA decomposition subroutine from Python's scikit-learn package\footnote{\url{https://scikit-learn.org/stable/modules/generated/sklearn.decomposition.PCA.html}}.

\begin{table}[!h]
    \begin{center}
    \caption{Variance attributed to each Principal Component axis}\label{tab:PCA_var}
    \begin{tabular}{ccc}
    \tableline
    Principal Component & ASAS-SN & WISE  \\
    \tableline
    \tableline
    Axis 1 & 54.23\% & 56.51\%  \\
    Axis 2 & 38.67\% & 35.40\%  \\
    Axis 3 & 7.10\% & 8.09\% \\
    \tableline
    \end{tabular} 
    \end{center}
\end{table}

Table~\ref{tab:PCA_var} shows the PCA variance for each of the ASAS-SN and WISE datasets. It is clear that both samples can be effectively described with just the first two axes, with very little scatter associated to the third component, orthogonal to the plane described by axis 1 and 2. This shows that our data can be efficiently represented by a plane. For both datasets, the first axis is well aligned with the period of the stars, while the second axis is an almost equally weighted combination of $\phi_{31}$ and [Fe/H]. Based on this result, we feel confident that our dataset can be represented by a linear fit of period, $\phi_{31}$, and [Fe/H]. We determine the exact coefficients using the ODR method described in Section~\ref{ssec:planefit}. This method still provides rotated axes for the fit, but it allows us to do so while including uncertainties in the fitting variables and data.

\section{K-Nearest Neighbor Analysis} \label{sec:KNN}

In this work, we chose to represent the relation between [Fe/H], period, and $\phi_{31}$ of RRab stars using a linear function of these three variables (see Section~\ref{ssec:planefit}). An alternative approach consists of adopting non-parametric techniques that could provide a similar result without assumptions about the functional form of the relation to fit. One such technique is the $k$-Nearest Neighbors ($k$-NN) method \citep{1053964}.

While traditionally used in machine learning classification problems (see e.g. \citealt{2009AJ....138...63M} for an application in astronomy), the $k$-NN method also serves as a powerful non-parametric regression technique to estimate the value of a given variable based on the values of the closest neighbors, in a properly defined $n$-dimensional space. In this work, the location on the period-$\phi_{31}$ plane is used as a predictor for [Fe/H]. Following \citet{2007ApJ...663..774B}, we calculate the Euclidean distance, in the period-$\phi_{31}$ plane, of each star from  every other star in the sample. We then estimate the $k$-NN metallicity of each star as the [Fe/H] weighted average of its nearest $k$ neighbors, where $k$ is a suitable integer number. The weights used are proportional to the inverse distance in the period-$\phi_{31}$ plane from the test star so that nearer stars contribute more to the average than the farthest sources.

\begin{figure}[!h]
    \centering
    \includegraphics[width=\textwidth]{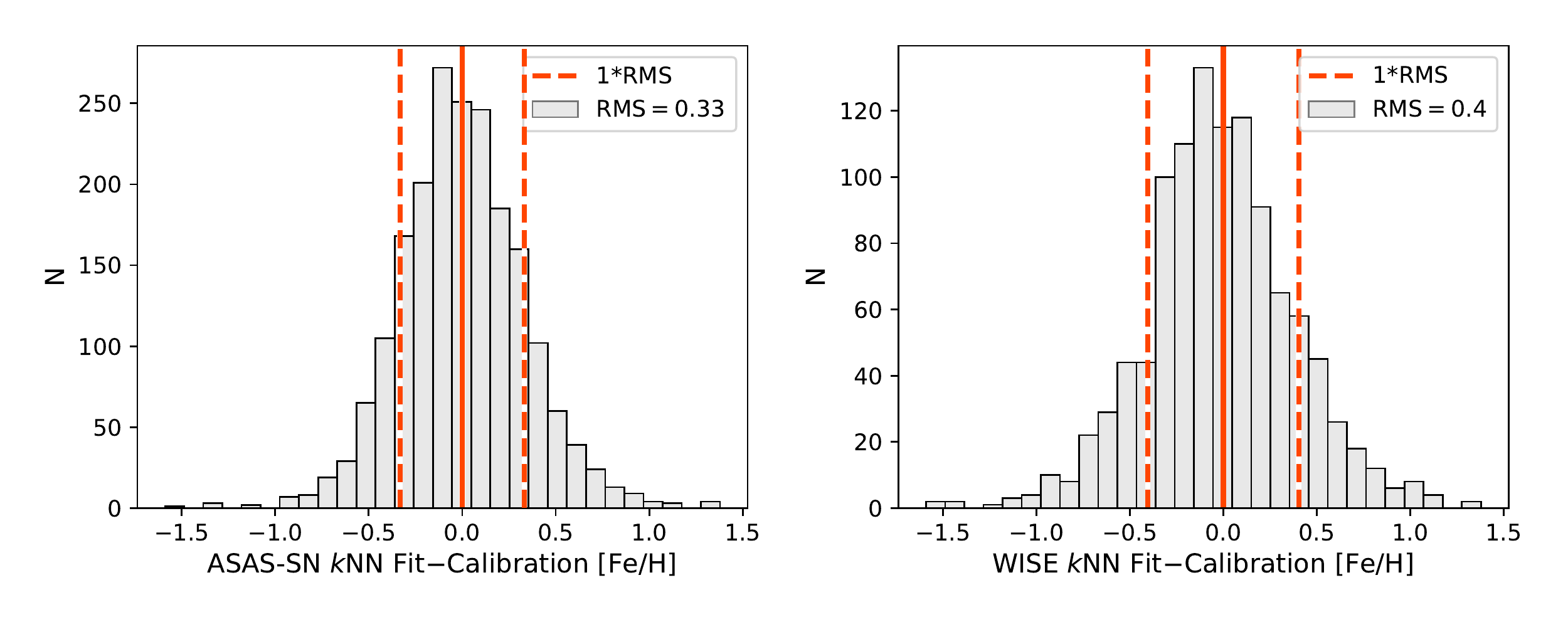}
    \caption{Difference in the [Fe/H] predicted by the $k$-NN method to the metallicity of the calibration sample for ASAS-SN dataset (left) and WISE dataset (right). The root mean square error between the $k$-NN and calibration metallicity is listed in the top right corners and shown by the vertical dashed lines.}
    \label{fig:KNN}
\end{figure}

The number of nearest neighbors $k$ is chosen in order to balance the need of averaging-out the natural scatter in the neighbor’s [Fe/H], while at the same time still preserving predictive power at the sparse edges of the samples' distribution in the period-$\phi_{31}$ plane. Since our goal is to capture the global trend over the entire range of values, and not only the highest density region, we chose $k$ = 5 as the optimum number of near neighbors for both our samples.

Figure~\ref{fig:KNN} shows the difference between the metallicity derived with the $k$-NN method and the spectroscopic [Fe/H] for our optical and infrared samples. The Root Mean Square (RMS) [Fe/H] error of the residuals is 0.33~dex and 0.40~dex for the ASAS-SN and WISE samples, respectively. Since the $k$-NN method, in contrast to parametric fits, does not rely upon any particular functional form, we can assume that the quoted RMS errors are representative of the best possible overall [Fe/H] uncertainty (averaged over the entire range of metallicity) that a specific fit could achieve with the available data.



\end{document}